\documentclass[a4paper]{article}
\usepackage[utf8]{inputenc}
\usepackage[T1]{fontenc}
\usepackage{geometry}
\usepackage{array}
\usepackage{booktabs}
\usepackage{longtable}
\usepackage{hyperref}
\usepackage{xcolor}
\usepackage{listings}
\usepackage{amsmath}
\usepackage{algorithmicx}
\usepackage{amssymb}
\usepackage{graphicx}
\usepackage{tabularx}
\usepackage{algorithm}
\usepackage{algpseudocode}
\usepackage{comment}
\usepackage{float}
\usepackage{threeparttable}
\usepackage{authblk}
\usepackage{float}

\usepackage{tikz}
\usepackage{pgfplots} \usetikzlibrary{positioning,arrows.meta,decorations.pathreplacing,fit,calc,matrix}
\pgfplotsset{compat=1.18}
\hypersetup{
    colorlinks=true,
    linkcolor=blue,
    urlcolor=blue
}

\lstdefinestyle{bashstyle}{
    basicstyle=\ttfamily\small,
    breaklines=true,
    frame=single,
    backgroundcolor=\color{gray!8},
    columns=fullflexible
}

\title{Implementation and Optimization of HQC Decoding on NPU-Integrated Devices}

\author[1]{Vu Minh Chau}
\author[1]{Nguyen Ngoc Kiet}
\author[1]{Pham Quang Minh}
\author[1]{Mai Xuan Ngoc}
\author[1]{Nguyen Duc Anh}
\author[1]{Hoang Ta\thanks{Corresponding author}}

\affil[1]{ \small School of Information and Communication Technology, Hanoi University of Science and Technology, Vietnam}

\date{}

\begin{document}

\maketitle

\begin{abstract}
    Hamming Quasi-Cyclic (HQC) has been selected by NIST for standardization as an additional code-based key-encapsulation mechanism, providing algorithmic diversity alongside lattice-based post-quantum cryptography. Efficient deployment of HQC on mobile and embedded platforms, however, requires careful optimization of its decoding procedure, whose Reed-Muller and Reed-Solomon components dominate the computational cost. This paper studies HQC decoding on Qualcomm Hexagon processors in NPU-integrated devices, focusing on the Hexagon Vector eXtensions (HVX) backend rather than a tensor-inference engine. We observe that HQC decoding naturally exposes vector-structured computation, including Reed-Muller reliability vectors, Hadamard-transform coefficients, Reed-Solomon syndrome vectors, finite-field products, and packed support-point evaluations. Based on this observation, we redesign the dominant decoding kernels around HVX-friendly data layouts and execution patterns, including a vectorized Reed-Muller Hadamard transform, scalar-equivalent peak selection, HVX-oriented finite-field arithmetic, vectorized syndrome computation, and shortened-support locator-root evaluation. We implement and evaluate the optimized decoder using both Hexagon simulator measurements and real-device experiments on a Snapdragon~8 Gen~2 hardware development kit. The results show that Hexagon/HVX-assisted decoding substantially reduces latency and energy consumption, improving energy efficiency by up to $18.13\times$ while significantly offloading host CPU work. These results indicate that NPU-integrated mobile platforms can serve as effective backends for structured post-quantum cryptographic decoding when the underlying kernels are reformulated around vector execution.   
\end{abstract}
\section{Introduction}

Post-quantum cryptography has been widely studied in recent years, since classical public-key cryptographic primitives based on integer factorization and discrete logarithms are vulnerable to large-scale quantum attacks~\cite{shor1999polynomial}. In response to this long-term security concern, the U.S. National Institute of Standards and Technology (NIST) initiated a process to standardize quantum-resistant public-key cryptographic algorithms.
\par
Several families of cryptographic assumptions have been investigated as candidates for post-quantum security, including lattice-based, code-based, hash-based, multivariate-based, and isogeny-based assumptions. Among these, lattice-based cryptography has received particularly extensive attention and cryptanalytic study from the cryptographic community. As a result, two lattice-based constructions, namely the Module-Lattice-Based Digital Signature Algorithm (ML-DSA) and the Module-Lattice-Based Key-Encapsulation Mechanism (ML-KEM)~\cite{NIST24b,NIST24a}, were selected for standardization. In parallel, hash-based cryptography has also made substantial progress, with the Stateless Hash-Based Digital Signature Algorithm (SLH-DSA)~\cite{NIST24c} selected as a standardized stateless hash-based digital signature scheme. Beyond lattice-based and hash-based schemes, other post-quantum directions have also seen remarkable progress. In particular, code-based cryptography has recently gained further standardization momentum, with Hamming Quasi-Cyclic (HQC)~\cite{NIST25,HQCSpec2025} selected by NIST for standardization as an additional code-based key-encapsulation mechanism to complement and diversify the key-establishment portfolio alongside ML-KEM. Meanwhile, multivariate-based and isogeny-based signatures~\cite{NIST26SigRound3,Beullens21MAYO,DeFeoKLPW20SQISign} continue to be evaluated in NIST's additional digital signature standardization process. These efforts emphasize the importance and urgency of cryptographic diversity, mitigating the risk of a single point of failure.

\par
HQC~\cite{HQCSpec2025} relies on error-correcting codes and provides algorithmic diversity compared with lattice-based schemes. In essence, it employs a fixed concatenated code consisting of an outer Reed-Solomon code and an inner duplicated Reed-Muller code for error correction. As HQC has advanced toward standardization, recent studies have increasingly focused on its practical implementation, including software optimization, efficient finite-field arithmetic, and deployment on constrained or hardware-accelerated platforms~\cite{NIST25,DongFW25OptHQC,KimCSY25HQC,DeshpandeXNNS23HQC}. These works indicate that HQC performance is strongly influenced by how its decoding kernels are mapped to the target execution backend.
\par 
Modern mobile and edge platforms increasingly integrate heterogeneous accelerators to support computationally intensive data-parallel workloads, including signal-processing and neural-network inference workloads~\cite{ChenXieSCT20DNNAccel}. These accelerator resources are commonly exposed as part of Neural Processing Unit (NPU) subsystems, which combine scalar, vector, and tensor-style execution units~\cite{QualcommHexagonNPU}. On Qualcomm platforms, the Hexagon processor provides Hexagon Vector eXtensions (HVX), which support wide SIMD-style vector operations and enable efficient lane-wise arithmetic, comparisons, permutations, and reductions~\cite{Qua25}. Although HVX is primarily designed for signal-processing and AI-oriented workloads, its vector execution model is also well suited to structured cryptographic kernels, especially those underlying code-based primitives.

\par
In this work, we study HQC decoding on Qualcomm Hexagon processors in NPU-integrated devices. More specifically, our optimized kernels target the Hexagon/HVX vector backend rather than a tensor-inference engine. HQC decoding is a natural candidate for this execution model because its dominant Reed-Muller and Reed-Solomon kernels operate on vector-structured data, including Reed-Muller reliability vectors, Hadamard-transform coefficients, Reed-Solomon syndrome vectors, and packed finite-field support points. At the same time, these kernels cannot be accelerated efficiently by a direct translation of the scalar decoder: the Hadamard transform requires HVX-friendly data rearrangement, peak selection must preserve the scalar tie-breaking rule, and Reed-Solomon decoding requires vectorized finite-field arithmetic and support-point evaluation. We therefore decompose the HQC decoder into its Reed-Muller and Reed-Solomon components and redesign the dominant decoding kernels around HVX-friendly data layouts and execution patterns. The goal is not only to reduce processor-cycle counts, but also to improve energy efficiency and offload host CPU work on mobile and embedded platforms. 

\subsection*{Contributions}

Our contributions are as follows.

\begin{itemize}
\item We identify HQC decoding as a natural candidate for Hexagon/HVX acceleration on NPU-integrated mobile platforms. Its dominant Reed-Muller and Reed-Solomon decoding kernels operate on vector-structured data, including reliability vectors, Hadamard-transform coefficients, syndrome vectors, finite-field elements, and packed support points. Based on this observation, we present an end-to-end optimized HQC decoding implementation targeting the Hexagon/HVX vector backend.

\item We redesign the dominant decoding kernels around HVX-friendly data layouts and execution patterns. For the inner duplicated Reed-Muller code, we develop an HVX-friendly fast Hadamard transform and a vectorized peak-selection procedure that exactly preserves the scalar tie-breaking rule. For the outer Reed-Solomon stage, we introduce HVX-oriented finite-field multiplication, vectorized syndrome computation, and a shortened-support Chien search, namely a vectorized evaluation of the error-locator polynomial over the public shortened Reed-Solomon support points. In selected substage benchmarks, these optimizations reduce the Reed-Muller Hadamard transform from $263{,}175$ to $17{,}950$ Pcycles per decode, peak selection from $71{,}217$ to $6{,}081$ Pcycles, syndrome computation from $162{,}312$ to $3{,}517$ Pcycles, and the error-locator-polynomial-related stage from $119{,}595$ to $6{,}581$ Pcycles.

\item We evaluate the optimized decoder using both Hexagon simulator measurements and real-device experiments. In simulator measurements, the optimized implementation reduces the full HQC-128 decoding cost from $953{,}763$ to $41{,}471$ Pcycles per decode, corresponding to a $23.00\times$ speedup and a $95.7\%$ reduction in Pcycles. On a Snapdragon~8 Gen~2 hardware development kit, the Hexagon/HVX-assisted implementation achieves $2.07\times$, $1.85\times$, and $1.96\times$ latency speedups for HQC-128, HQC-192, and HQC-256, respectively; improves
energy efficiency by $18.13\times$, $11.77\times$, and $16.81\times$; and reduces host CPU ms/decode by $99.0\%$--$99.7\%$ across all parameter sets.
\end{itemize}

\subsection*{Related Work}

\paragraph{HQC implementations on different hardware backends.}
Recent work has explored HQC implementation across a range of execution environments. On embedded CPUs, optimized HQC implementations on ARM Cortex-M4 platforms have been proposed to reduce the cost of polynomial multiplication and finite-field operations~\cite{KimCSY25HQC,Aissaoui24HQC}. On reconfigurable hardware, several works study HQC acceleration using FPGA and RTL designs. Deshpande et al. present a hand-optimized Verilog implementation of HQC key generation, encapsulation, and decapsulation~\cite{DeshpandeXNNS23HQC}, while later RTL accelerators provide unified support for HQC-128, HQC-192, and HQC-256~\cite{AntognazzaBPS25HQC}. Other works target specific bottlenecks, such as sparse polynomial multiplication or Frobenius additive FFT-based multiplication on RISC-V/FPGA SoC platforms~\cite{TuHKX23LEAP,RasLCPSRV25HQC}. These studies show that HQC performance strongly depends on the execution backend and the mapping of its algebraic kernels to hardware-friendly operations.

\paragraph{Hardware acceleration of other post-quantum primitives.}
Beyond HQC, hardware-aware optimization has also been widely studied for other post-quantum cryptographic schemes. For lattice-based schemes, several works optimize Kyber/ML-KEM and Dilithium/ML-DSA on ARM Cortex-M4 by accelerating their number-theoretic transform and modular arithmetic kernels~\cite{AbdulrahmanHKS22KyberDilithium}. FPGA-based accelerators have also been developed for high-volume Kyber and Dilithium workloads, using parallelism and batch processing to improve throughput~\cite{Carril24HighVolumePQC}. More recent hardware/software co-design efforts target open-source silicon platforms such as OpenTitan, where specialized big-number accelerators are used to speed up NTT-based PQC primitives~\cite{UrquhartS24OpenTitan}. Compared with these CPU, microcontroller, FPGA, and RISC-V/SoC implementations, our work focuses on NPU-integrated Qualcomm platforms and studies how HQC decoding can be mapped to Hexagon/HVX vector operations to improve execution time, energy efficiency, and CPU offload.

\paragraph{Organization.}
The remainder of this paper is organized as follows. Section~\ref{sec:preliminaries} reviews the fundamental background on first-order Reed-Muller codes, Reed-Solomon codes, the HQC concatenated-code structure, and the NPU/HVX execution model used throughout the paper. Section~\ref{sec:proposed} presents the proposed HQC decoding optimizations. Section~\ref{sec:experiment} evaluates the proposed implementation using both Hexagon simulator measurements and real-device experiments, reporting latency, energy efficiency, and CPU-offloading benefits. Section~\ref{sec:conclusion} concludes the paper. Finally, Section~\ref{sec:appendix} provides supplementary implementation details, including pseudocode for the main optimized kernels and the mapping between the abstract vector operations and Qualcomm Hexagon HVX intrinsics.

\section{Preliminaries}
\label{sec:preliminaries}

We use the following notation throughout the paper. Let $\mathbb{Z}$ denote the
ring of integers, and let $\mathbb{F}_2$ denote the binary finite field. For
$q=2^8$, we write $\mathbb{F}_q$ for the finite field used by the
Reed-Solomon component of HQC. For a positive integer $m$, let $[m]=\{1,\ldots,m\}$. Vectors are denoted in boldface, e.g., $\mathbf{x}$, and $x_i$ denotes the $i$-th coordinate of $\mathbf{x}$. 

\subsection{First-Order Reed-Muller Code}





Let $m$ be a non-negative integer. The first-order Reed-Muller code of length $2^m$, denoted by $\mathrm{RM}(1,m)$, is defined as the set of evaluation vectors of all affine Boolean functions over $\mathbb{F}_2^m$. More precisely,
\[
    \mathrm{RM}(1,m)
    =
    \left\{
    \bigl(f(\mathbf{x})\bigr)_{\mathbf{x}\in\mathbb{F}_2^m}
    :
    f(\mathbf{x})=b+\sum_{i\in[m]}a_i x_i \pmod 2
    \right\}
    \subseteq \mathbb{F}_2^{2^m}.
\]
The parameters of $\mathrm{RM}(1,m)$ are $n=2^m,\, k=m+1,\, d=2^{m-1},$
where $n$ is the code length, $k$ is the dimension, and $d$ is the minimum Hamming distance. Decoding first-order Reed-Muller codes can be performed efficiently using the fast Hadamard transform. Given a received vector $\mathbf{y}=(y_{\mathbf{x}})_{\mathbf{x}\in\mathbb{F}_2^m}$, the decoder first
maps its binary entries to the bipolar representation $z_{\mathbf{x}}=(-1)^{y_{\mathbf{x}}}$. For each
$\mathbf{a}\in\mathbb{F}_2^m$, it computes the correlation
\[
    W(\mathbf{a})
    =
    \sum_{\mathbf{x}\in\mathbb{F}_2^m}
    z_{\mathbf{x}}(-1)^{\langle \mathbf{a},\mathbf{x}\rangle}.
\]
The decoder selects a value of $\mathbf{a}$ that maximizes $|W(\mathbf{a})|$. The sign of $W(\mathbf{a})$ determines the affine offset, and the corresponding affine function is returned as the decoded codeword.

 \subsection{Reed-Solomon Code}



Let $\mathcal{R}=\mathbb{F}_q$ and let $\mathcal{S}=\{\alpha_1,\ldots,\alpha_n\}\subseteq\mathcal{R}$ be a set of distinct evaluation points with $n\le q$. For a message polynomial $m(X)\in\mathbb{F}_q[X]$ satisfying $\deg(m)<k$, the Reed-Solomon code of
length $n$ and dimension $k$ is defined as
\[
    \mathsf{RS}(n,k)
    =
    \left\{
    \bigl(m(\alpha_1),\ldots,m(\alpha_n)\bigr)
    :
    m(X)\in\mathbb{F}_q[X],\ \deg(m)<k
    \right\}.
\]
Its minimum Hamming distance is
\[
    d=n-k+1.
\]

For decoding, let $\mathbf{r}=\mathbf{c}+\mathbf{e}$ be the received word, where $\mathbf{c}$ is the transmitted codeword and $\mathbf{e}$ is the error vector. The decoder first computes the syndrome values
\[
    S_j = r(\alpha^j), \qquad j=1,\ldots,2\delta,
\]
where $\alpha$ is a primitive element of $\mathbb{F}_q$ and $\delta$ is the error-correcting capability. In the cyclic or parity-check formulation used by the decoder, every valid codeword has zero syndrome at these check points. Hence, the syndromes depend only on the error polynomial $e(X)$. If all syndromes are zero, then $\mathbf{r}$ is already a valid codeword.


Otherwise, the decoder uses the syndrome sequence to solve the key equation and recover the error-locator polynomial
\[
    \Lambda(X)=\prod_{i\in\mathcal{E}}(1-X\alpha^i),
\]
where $\mathcal{E}$ is the set of error positions. The roots of $\Lambda(X)$ identify the error locations, and the corresponding error values are then
recovered from the error-evaluator polynomial. Finally, these errors are subtracted from $\mathbf{r}$ to obtain the original codeword.

Although Reed-Solomon codes are introduced above in the evaluation-code form, the HQC decoder uses a shortened cyclic/parity-check representation. In this representation, the received word is viewed as a polynomial \(r(X)=\sum_{i=0}^{n_1-1} r_iX^i\), and syndrome values are computed by evaluating this polynomial at the prescribed check points. Thus, valid codewords have zero syndromes at these check points.



\subsection{Hamming Quasi-Cyclic (HQC)}

HQC relies on a concatenated-code construction in which the inner and outer codes are Reed-Muller and Reed-Solomon codes, respectively. More precisely,
HQC uses an outer code with parameters $[n_e,k_e,d_e]$ over $\mathbb{F}_q$ and an inner binary code with parameters $[n_i,k_i,d_i]$ over $\mathbb{F}_2$, where $q=2^{k_i}$. Each symbol of $\mathbb{F}_q$ is mapped bijectively to a codeword of
the inner code, which induces a binary mapping
\[
    \mathbb{F}_q^{n_e} \longrightarrow \mathbb{F}_2^N,
\]
where $N=n_e n_i$. Thus, the resulting binary concatenated code has parameters
\[
    [N=n_e n_i,\ K=k_e k_i,\ D\ge d_e d_i].
\]

HQC uses shortened Reed--Solomon codes together with duplicated Reed-Muller codes. The Reed-Muller parameters used in the three HQC parameter sets are
summarized in Table~\ref{tab:duplicated-rm-codes}.

\begin{table}[h]
\centering
\renewcommand{\arraystretch}{1.15}
\caption{Duplicated Reed--Muller codes.}
\label{tab:duplicated-rm-codes}
\begin{tabular}{|c|c|c|c|}
\hline
Instance & Reed--Muller code & Multiplicity & Duplicated Reed--Muller code \\
\hline
HQC-128 & $[128,8,64]$ & $3$ & $[384,8,192]$ \\
\hline
HQC-192 & $[128,8,64]$ & $5$ & $[640,8,320]$ \\
\hline
HQC-256 & $[128,8,64]$ & $5$ & $[640,8,320]$ \\
\hline
\end{tabular}
\end{table}

\noindent
The multiplicity indicates how many times each Reed-Muller codeword is duplicated.


\subsection{Neural Processing Unit (NPU)}

In this work, the term refers to the data-parallel execution model exposed by vector accelerators such as Qualcomm Hexagon HVX~\cite{Qua25}. Rather than describing the optimized decoder solely in terms of processor-specific instructions, we model the accelerator as a finite-lane vector machine. This abstraction is sufficient for the Reed-Muller and Reed-Solomon optimizations considered in this work, since their dominant operations consist of regular lane-wise arithmetic, finite-field products, permutations, and reductions.

We mainly focus on the HQC-128 parameter set. The same optimization principles also apply to the other HQC parameter sets, with the corresponding changes in block sizes and the number of required vector blocks. Let $L$ denote the number of lanes in a vector register, and let
\[
    \mathbf{x}=(x_0,\ldots,x_{L-1})\in\mathcal{R}^{L}.
\]
The ring $\mathcal{R}$ depends on the stage being accelerated. Signed 16-bit integers are used for Reed-Muller soft values, whereas elements of
$\mathbb{F}_{2^8}$ are embedded in 16-bit lanes for Reed-Solomon arithmetic. We use the following abstract vector operations:
\[
\begin{aligned}
\mathsf{VADD}(\mathbf{x},\mathbf{y}) &=(x_i+y_i)_{i=0}^{L-1},\\
\mathsf{VSUB}(\mathbf{x},\mathbf{y}) &=(x_i-y_i)_{i=0}^{L-1},\\
\mathsf{VABS}(\mathbf{x}) &=(|x_i|)_{i=0}^{L-1},\\
\mathsf{VMAX}(\mathbf{x},\mathbf{y}) &=(\max(x_i,y_i))_{i=0}^{L-1},\\
\mathsf{VMIN}(\mathbf{x},\mathbf{y}) &=(\min(x_i,y_i))_{i=0}^{L-1}.
\end{aligned}
\]
The operation
\[
    \mathsf{VSPLAT}(a)=(a,\ldots,a)
\]
broadcasts a scalar value to all vector lanes. We represent bit-level vector operations as
\[
\begin{aligned}
\mathsf{VXOR}(\mathbf{x},\mathbf{y}) &=(x_i\oplus y_i)_{i=0}^{L-1},\\
\mathsf{VAND}(\mathbf{x},\mathbf{y}) &=(x_i\wedge y_i)_{i=0}^{L-1},\\
\mathsf{VSHL}_{s}(\mathbf{x}) &=(x_i\ll s)_{i=0}^{L-1},\\
\mathsf{VSHR}_{s}(\mathbf{x}) &=(x_i\gg s)_{i=0}^{L-1}.
\end{aligned}
\]

We also use two reduction operations:
\[
    \mathsf{VREDUCE\_MAX}(\mathbf{x})=\max_{0\le i<L}x_i,
    \qquad
    \mathsf{VREDUCE\_MIN}(\mathbf{x})=\min_{0\le i<L}x_i.
\]
Both reductions are implemented as rotation trees. For example, maximum reduction applies updates of the form
\[
    \mathbf{z}
    \leftarrow
    \mathsf{VMAX}
    \bigl(
        \mathbf{z},
        \mathsf{VROT}_{s}(\mathbf{z})
    \bigr)
\]
for a sequence of rotation offsets $s$. Replacing $\mathsf{VMAX}$ with $\mathsf{VMIN}$ gives the corresponding minimum reduction. Finally, the Reed--Muller transform uses the following selection primitives:
\[
\begin{aligned}
\mathsf{VCMPEQ}(\mathbf{x},\mathbf{y})
    &=
    \bigl(\mathbf{1}\{x_i=y_i\}\bigr)_{i=0}^{L-1},\\
\mathsf{VSELECT}(\mathbf{m},\mathbf{x},\mathbf{y})
    &=
    (m_i x_i+(1-m_i)y_i)_{i=0}^{L-1}.
\end{aligned}
\]
The complete mapping between the abstract vector operations and the corresponding Qualcomm Hexagon HVX intrinsics is provided in Section~\ref{tab:hvx_mapping}.

\section{Proposed Algorithms}
\label{sec:proposed}


\subsection{Motivation and Challenges}

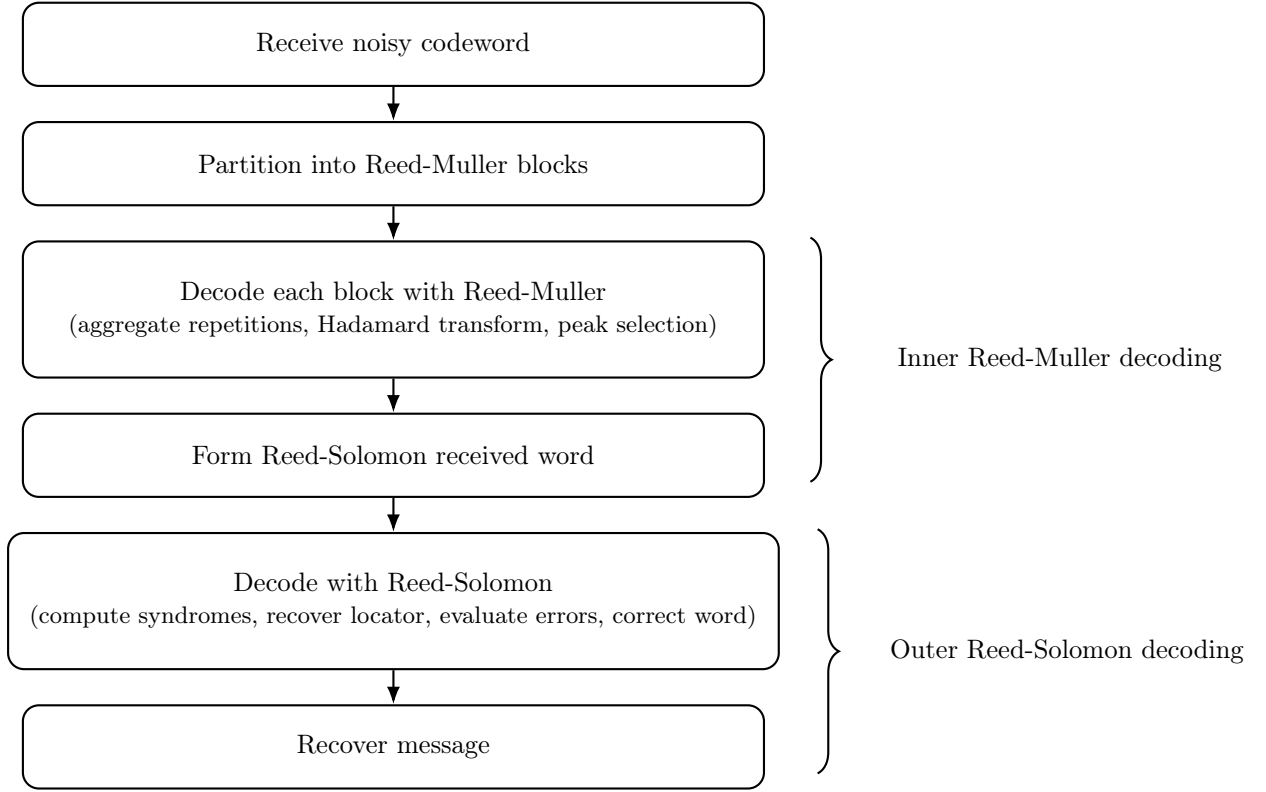
\begin{figure}[h]
\centering
\begin{tikzpicture}[
    node distance=0.45cm,
    >=Latex,
    box/.style={
        draw,
        rounded corners=6pt,
        thick,
        minimum width=9.8cm,
        minimum height=1.1cm,
        align=center,
        inner sep=6pt
    },
    bigbox/.style={
        draw,
        rounded corners=6pt,
        thick,
        minimum width=9.8cm,
        minimum height=1.8cm,
        align=center,
        inner sep=8pt
    },
    arrow/.style={->, thick}
]

\node[box] (recv) {Receive noisy codeword};

\node[box, below=of recv] (part) {Partition into Reed-Muller blocks};

\node[bigbox, below=of part] (rmdec) {%
Decode each block with Reed-Muller\\
\small (aggregate repetitions, Hadamard transform, peak selection)
};

\node[box, below=of rmdec] (rsword) {Form Reed-Solomon received word};

\node[bigbox, below=of rsword] (rsdec) {%
Decode with Reed-Solomon\\
\small (compute syndromes, recover locator, evaluate errors, correct word)
};

\node[box, below=of rsdec] (msg) {Recover message};

\draw[arrow] (recv) -- (part);
\draw[arrow] (part) -- (rmdec);
\draw[arrow] (rmdec) -- (rsword);
\draw[arrow] (rsword) -- (rsdec);
\draw[arrow] (rsdec) -- (msg);

\draw[decorate, decoration={brace, amplitude=8pt}, thick]
    ($(rmdec.east)+(0.6cm,0.95cm)$)
    -- ($(rmdec.east |- rsword.east)+(0.6cm,-0.35cm)$)
    node[midway, xshift=3.3cm] {Inner Reed-Muller decoding};

\draw[decorate, decoration={brace, amplitude=8pt}, thick]
    ($(rsdec.east)+(0.5cm,0.95cm)$)
    -- ($(rsdec.east |- msg.east)+(0.5cm,-0.35cm)$)
    node[midway, xshift=3.3cm] {Outer Reed-Solomon decoding};

\end{tikzpicture}
\caption{Overview of the HQC decoding process.}
\label{fig:hqc-decoding-overview}

\end{figure}

HQC decoding is a natural candidate for acceleration on the Hexagon/HVX vector backend because its main computational objects have an inherent vector structure. In the Reed-Muller component, each duplicated block is first converted into a reliability vector and then processed by a Hadamard transform whose butterfly stages consist of regular additions, subtractions, and data rearrangements. The summary of HQC decoding components is illustrated in Figure~\ref{fig:hqc-decoding-overview}. In the Reed-Solomon component, syndrome values, finite-field products, and evaluations over the shortened support can also be organized as packed vectors. These structures match the lane-wise execution model of HVX and make it possible to exploit data-level parallelism beyond a scalar CPU implementation.
\par
At the same time, HQC decoding cannot be efficiently accelerated by a direct translation of the scalar decoder. For Reed-Muller decoding, the Hadamard transform requires an HVX-friendly memory layout and a careful sequence of vector rearrangements. Moreover, the final peak-selection step must preserve the scalar decoder's tie-breaking rule, since different choices among equal-magnitude Hadamard coefficients may lead to different decoded outputs. Therefore, vectorization must preserve not only performance but also bit-level equivalence with the scalar reference behavior.
\par
The Reed-Solomon stage introduces a different set of challenges. Finite-field multiplication over \(\mathbb{F}_{2^8}\), syndrome computation, locator-root evaluation, and error-locator-polynomial updates involve a mixture of regular vector operations and scalar control logic. In particular, the additive-FFT-style root evaluation used in scalar implementations contains sequential dependencies and irregular memory-access patterns that are not well aligned with HVX execution. We therefore redesign the dominant Reed-Solomon kernels around HVX-oriented scalar-by-vector finite-field multiplication, vectorized syndrome computation and shortened-support Chien search.
\par
We describe the optimized implementation for the HQC-128 parameter set. The same optimization principles extend to HQC-192 and HQC-256,
with changes in block sizes and the number of required vector blocks.

\paragraph{Decoding vs.\ Full Decapsulation.} HQC decapsulation proceeds in two stages: it first recovers a noisy codeword by combining the ciphertext with the secret key through a sparse polynomial multiplication, and then applies the concatenated decoder to that noisy codeword. This paper optimizes only the second stage---the decoding of the noisy codeword, i.e., the entry point of the pipeline in Figure~\ref{fig:hqc-decoding-overview}. The
polynomial-multiplication stage that produces the noisy codeword is a separate and comparably significant cost in HQC implementations, and has been the subject of dedicated accelerators, including sparse-polynomial multipliers~\cite{TuHKX23LEAP} and Frobenius additive-FFT multipliers~\cite{RasLCPSRV25HQC}; it is orthogonal to the decoding kernels considered here. Accordingly, the speedups reported in this paper are for the decoding stage in isolation, not for end-to-end decapsulation. We focus on decoding because its Reed-Muller and Reed-Solomon kernels expose the non-obvious, control-flow-sensitive vectorization challenges (Hadamard data layout, scalar-equivalent tie-breaking, and locator-root evaluation), whereas the sparse polynomial multiplication is a regular shift-and-XOR pattern that maps directly onto the same HVX backend; integrating it is a natural next step toward full-decapsulation acceleration.

\subsection{Reed-Muller Decoding}

\subsubsection{Repetition aggregation.} For each Reed-Solomon symbol, the duplicated Reed-Muller code provides several repetitions of a length-$128$ Reed-Muller codeword. In HQC-128, the multiplicity is three. The decoder first aggregates the repeated bits coordinate-wise. If $b_i^{(r)}\in\{0,1\}$ denotes the $i$-th bit in repetition $r$, then the soft value used by the Hadamard decoder is
\[
    x_i=\sum_{r=0}^{2} b_i^{(r)}, \qquad 0\le i<128.
\]
This produces an integer reliability vector that is subsequently used as the input to the Hadamard transform. For HQC-192 and HQC-256, the same aggregation
procedure is used with multiplicity five.

\subsubsection{Vector Hadamard transformation.} After expansion, the Reed-Muller decoder applies the fast Hadamard transform
for \(\mathrm{RM}(1,7)\). The current vector is split into two
paired parts
$
    \mathbf{x} = \mathbf{x}_{L} \Vert \mathbf{x}_{R},
$
 then the decoder computes the lane-wise sum and difference,
\[
    \mathbf{y}_{L} = \mathbf{x}_{L} + \mathbf{x}_{R},
    \qquad
    \mathbf{y}_{R} = \mathbf{x}_{L} - \mathbf{x}_{R},
\]
and concatenates them to form the next transform state
$
    \mathbf{y} = \mathbf{y}_{L} \Vert \mathbf{y}_{R}
$ (Figure~\ref{fig:consecutive-blocks}).
We utilize the $\mathsf{VDEALH}$ intrinsic to split vector $\mathbf{x}$ before computing element-wise addition and subtraction in parallel using $\mathsf{VADDH}$ and $\mathsf{VSUBH}$, respectively. The transform consists of seven butterfly stages for the $128$-point Hadamard
transform. The complete procedure is given in Algorithm~\ref{alg:hadamard-hvx}.

\usetikzlibrary{positioning,arrows.meta,calc}
\begin{figure}[h]
\centering
\begin{tikzpicture}[
    block/.style={
        draw,
        minimum width=1.0cm,
        minimum height=0.65cm,
        align=center
    },
    bluearrow/.style={->, thick, blue!70},
    redarrow/.style={->, thick, red!70}
]

\node[block,draw=blue!50,fill=blue!20] (b0) {$b_0$};
\node[block,right=0pt of b0,draw=blue!50,fill=blue!20] (b1) {$b_1$};
\node[block,right=0pt of b1] (dotsA) {$\cdots$};
\node[block,right=0pt of dotsA,draw=blue!50,fill=blue!20] (b63) {$b_{63}$};

\node[block,right=0pt of b63,draw=red!50,fill=red!20] (b64) {$b_{64}$};
\node[block,right=0pt of b64] (dotsB) {$\cdots$};
\node[block,right=0pt of dotsB,draw=red!50,fill=red!20] (b126) {$b_{126}$};
\node[block,right=0pt of b126,draw=red!50,fill=red!20] (b127) {$b_{127}$};

\node[block,below=of b0,xshift=-10mm,draw=blue!50,fill=blue!20] (b01) {$b_0$};
\node[block,right=0pt of b01,draw=blue!50,fill=blue!20] (b11) {$b_1$};
\node[block,right=0pt of b11] (dotsL) {$\cdots$};
\node[block,right=0pt of dotsL,draw=blue!50,fill=blue!20] (b631) {$b_{63}$};

\node[block,below=of b0,xshift=50mm,draw=red!50,fill=red!20] (b641) {$b_{64}$};
\node[block,right=0pt of b641] (dotsR) {$\cdots$};
\node[block,right=0pt of dotsR,draw=red!50,fill=red!20] (b1261) {$b_{126}$};
\node[block,right=0pt of b1261,draw=red!50,fill=red!20] (b1271) {$b_{127}$};

\node[font=\Large] at ($(b631.east)!0.5!(b641.west)$) {$\pm$};

\draw[bluearrow] (b0.south) -- (b01.north);
\draw[bluearrow] (b1.south) -- (b11.north);
\draw[bluearrow] (b63.south) -- (b631.north);

\draw[redarrow] (b64.south) -- (b641.north);
\draw[redarrow] (b126.south) -- (b1261.north);
\draw[redarrow] (b127.south) -- (b1271.north);

\end{tikzpicture}
\caption{Visual illustration of the butterfly operation with HVX support.}
\label{fig:consecutive-blocks}
\end{figure}
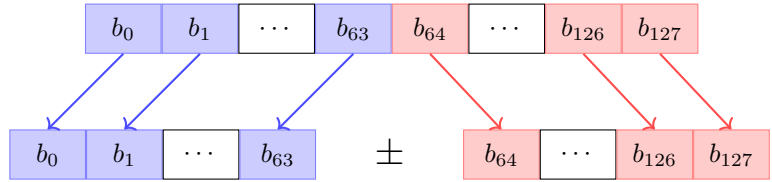

\subsubsection{Vectorized peak selection with scalar-equivalent tie breaking. } The final Reed-Muller decision selects the Hadamard coefficient with the largest
magnitude, breaking ties in favor of the smallest index:
\[
    i^\star = \min\arg\max_i |\hat{x}_i|.
\]

The HVX implementation follows the same rule, but performs the search in
parallel across all coefficient lanes. It first computes the magnitude vector
\[
    \mathbf m = (|\hat{x}_0|,\ldots,|\hat{x}_{127}|),
\]
and reduces it to the global maximum magnitude \(M\). All lanes satisfying
\(|\hat{x}_i|=M\) are then marked as peak candidates. To enforce the scalar
tie-breaking rule, the decoder keeps the original index of each candidate lane
and replaces every non-candidate lane by a sentinel value larger than any valid
index. Thus, the remaining problem becomes a vector minimum reduction over the
candidate indices. The result is exactly the smallest index attaining the
maximum magnitude.

Finally, the sign of the selected coefficient \(\hat{x}_{i^\star}\) determines the decoded Reed-Muller output bit. The detailed procedure is given in
Algorithm~\ref{alg:find-peaks-hvx}.

\subsection{Reed–Solomon Decoding}

For HQC-128, the shortened Reed-Solomon code has length \(n_1=46\), message
length \(k=16\), and error-correction parameter \(\delta=15\). 
Given a received word
$
    \mathbf{c}=(c_0,\ldots,c_{45})$,
the decoder computes \(2\delta=30\) syndromes, derives the error-locator polynomial \(\sigma(x)\), identifies its roots over the shortened support,
computes error magnitudes, and corrects the received word~\cite{HQCSpec2025}.

\subsubsection{Galois field multiplication.} Reed-Solomon decoding involves several polynomial operations over
\(\mathbb F_{2^8}\), where finite-field multiplication is one of the main costs. We use two multiplication strategies depending on the execution pattern:
table-driven multiplication for scalar operations, and an HVX-oriented bit-serial method when many products are computed in parallel.

\noindent\textbf{A. Table-driven multiplication.}
\label{sec:gf-mul}
For scalar products, we use logarithm and antilogarithm tables. The antilogarithm
table stores powers of the primitive element \(\alpha\), while the logarithm
table maps each nonzero field element to its exponent with respect to
\(\alpha\). Thus, instead of multiplying two field elements directly, we compute
\[
    a\cdot b =
    \begin{cases}
        0, & a=0 \text{ or } b=0,\\
        \alpha^{\log_\alpha(a)+\log_\alpha(b) \bmod 255}, & \text{otherwise}.
    \end{cases}
\]
This reduces one field multiplication to two logarithm lookups, one modular
addition, and one antilogarithm lookup.

\noindent\textbf{B. Vector multiplication.}
For vectorized Reed-Solomon operations, i.e., syndrome computation and Chien search, we
avoid table lookups and instead compute many field products in parallel using
HVX lane-wise operations. The target operation is scalar-by-vector
multiplication
\[
    a\cdot \mathbf b
    =
    (a b_0, a b_1,\ldots,a b_{L-1}),
    \qquad
    a,b_i\in \mathbb F_{2^8}.
\]

We first implement the lane-wise \(\mathsf{xtime}\) operation, i.e.,
multiplication by \(X\) in \(\mathbb F_{2^8}\). In HQC, an overflow after a left shift is reduced by XORing with
\(\mathtt{0x1d}\). Thus, for each byte \(x\),
\[
    \mathsf{xtime}(x)
    =
    \bigl((x \ll 1) \oplus (\mathsf{carry}\cdot \mathtt{0x1d})\bigr)
    \,\&\, \mathtt{0xff},
\]
where
$\mathsf{carry}=(x\gg 7)\,\&\,1.$
Using this \(\mathsf{xtime}\) primitive, scalar-by-vector multiplication is
performed in a bit-serial Horner form. Let $\mathbf{a}=(a,\dots,a)$, the bits of each lane \(b_i\) are then processed from the most significant bit
to the least significant bit. At bit position \(k\), the implementation extracts
the \(k\)-th bit of every lane,
\[
    \mathbf b_k=(\mathbf b\gg k)\,\&\,1,
\]
turns it into a full-lane mask \(\mathbf m_k\), updates the accumulator by
\(\mathsf{xtime}\), and conditionally XORs in \(\mathbf a\):
\[
    \mathbf z
    \leftarrow
    \mathsf{xtime}(\mathbf z)
    \oplus
    (\mathbf a\,\&\,\mathbf m_k).
\]
After the eight bit positions are processed, each lane of \(\mathbf z\)
contains \(a b_i\). The xtime and scalar vector multiplication procedures are given in Algorithm~\ref{alg:gf-xtime-hvx} and Algorithm~\ref{alg:gf-mul-scalar-vec-hvx}, respectively.
\subsubsection{Vectorized syndrome evaluation}

Syndrome computation evaluates the received Reed-Solomon word at consecutive powers of
the primitive element. Instead of computing one syndrome at a time, we vectorize
across syndrome indices.

For each received-symbol position \(j\), we pack the required powers into
\[
    \mathbf a_j =
    \bigl(
        \alpha^j,\alpha^{2j},\ldots,\alpha^{(2\delta)j},
        0,\ldots,0
    \bigr)
    \in \mathbb F_{2^8}^{64}.
\]
The first \(2\delta\) lanes correspond to syndrome values, while the remaining
lanes are padding.

The decoder invokes the pre-described vector multiplication for each received symbol $c_j$ and $\mathbf{a}_j$ to get the output:
\[
    c_j \mathbf a_j
    =
    (c_j\alpha^j,\;c_j\alpha^{2j},\ldots,c_j\alpha^{(2\delta)j},0,\ldots,0)
\]
and accumulates it by XOR:
\[
    \mathbf s \leftarrow \mathbf s \oplus c_j\mathbf a_j .
\]
After all \(j\ge 1\) terms are processed, the constant term \(c_0\) is
broadcast to all lanes and XORed into the accumulator:
\[
    \mathbf S = \mathbf s \oplus \mathsf{VSPLAT}(c_0).
\]
The first \(2\delta\) lanes of \(\mathbf S\) form the syndrome array.

\subsubsection{Shortened-support root search}

\label{sec:rs_code_3_3_3}
Root finding identifies the positions where the error-locator polynomial
vanishes on the Reed-Solomon support. For a candidate position \(i\), let
\[
    x_i = \alpha^{-i}, \qquad 0 \le i < n_1 .
\]

A position is marked as erroneous when the locator polynomial evaluates to zero at this point. The HQC specification describes this step using an additive FFT, which is well suited to scalar CPU implementations due to its asymptotically efficient polynomial evaluation structure. However, additive FFTs contain several sequential data dependencies and irregular memory-access patterns that limit their efficiency on HVX-style vector accelerators. Therefore, in our implementation, we instead use a shortened-support Chien search~\cite{Chien64}, whose evaluation pattern maps more naturally to lane-wise SIMD execution. In particular, we evaluate the locator polynomial only on the public support points of the shortened code.

To expose vector parallelism, the powers of all support points are packed into
vectors. For each degree \(j\), we define
\[
    \mathbf{p}_j
    =
    (x_0^j,x_1^j,\ldots,x_{n_1-1}^j,0,\ldots,0)
    \in \mathbb{F}_{2^8}^{64},
    \qquad
    0\le j\le \delta,
\]
where the remaining lanes are padded to match the HVX vector width. The locator
polynomial is then evaluated by accumulating the coefficient-weighted support
vectors, which are again computed efficiently with the help of scalar vector multiplication described above, using \texttt{Q6\_V\_vxor\_VV}:
\[
    \mathbf{r}
    =
    \sigma_0\mathbf{p}_0
    \oplus
    \sigma_1\mathbf{p}_1
    \oplus
    \cdots
    \oplus
    \sigma_{\deg\sigma}\mathbf{p}_{\deg\sigma}.
\]
Equivalently, the \(i\)-th active lane of \(\mathbf r\) contains
\(\sigma(x_i)\). Lanes that evaluate to zero give
the located-error indicator
\[
    e_i^{\mathrm{loc}} = \mathbf{1}\{\sigma(x_i)=0\},
    \qquad 0\le i<n_1.
\]

\begin{figure}[h]
\centering
\resizebox{0.92\linewidth}{!}{%
\begin{tikzpicture}[
    >=Latex,
    font=\scriptsize,
    coeff/.style={
        draw=green!50!black,
        fill=green!12,
        rounded corners=3pt,
        thick,
        minimum width=11mm,
        minimum height=5.5mm,
        align=center
    },
    acc/.style={
        draw=red!65!black,
        fill=red!8,
        rounded corners=4pt,
        thick,
        minimum width=8.8cm,
        minimum height=9mm,
        align=center,
        inner sep=4pt
    },
    comp/.style={
        draw=black,
        rounded corners=3pt,
        thick,
        minimum width=3.2cm,
        minimum height=7mm,
        align=center,
        inner sep=3pt
    },
    mask/.style={
        draw=orange!80!black,
        fill=orange!13,
        rounded corners=4pt,
        thick,
        minimum width=7.8cm,
        minimum height=7mm,
        align=center,
        inner sep=4pt
    },
    flow/.style={->, thick},
    plain/.style={thick}
]

\def\cellw{1.00}
\def\cellh{0.50}
\def\vecx{-2.45}
\def\busx{4.85}

\def\yzero{-0.85}
\def\yone{-1.90}
\def\yd{-3.35}

\coordinate (p0sw) at (\vecx,\yzero);
\foreach \i/\txt in {0/{1},1/{\cdots},2/{1},3/{0},4/{\cdots},5/{0}}{
    \pgfmathsetmacro{\xx}{\i*\cellw}
    \draw[draw=blue!65!black,fill=blue!8]
        ([xshift=\xx cm]p0sw) rectangle ++(\cellw,\cellh);
    \node at ([xshift=\xx cm + 0.5*\cellw cm,yshift=0.5*\cellh cm]p0sw)
        {$\txt$};
}
\node[above=0mm] at ($(p0sw)+(3.0,0.50)$)
    {$\mathbf p_0=(1,\ldots,1,0,\ldots,0)$};

\coordinate (p1sw) at (\vecx,\yone);
\foreach \i/\txt in {0/{x_0},1/{\cdots},2/{x_{n-1}},3/{0},4/{\cdots},5/{0}}{
    \pgfmathsetmacro{\xx}{\i*\cellw}
    \draw[draw=blue!65!black,fill=blue!8]
        ([xshift=\xx cm]p1sw) rectangle ++(\cellw,\cellh);
    \node at ([xshift=\xx cm + 0.5*\cellw cm,yshift=0.5*\cellh cm]p1sw)
        {$\txt$};
}
\node[above=0 mm] at ($(p1sw)+(3.0,0.50)$)
    {$\mathbf p_1=(x_0,\ldots,x_{n_1-1},0,\ldots,0)$};

\coordinate (pdsw) at (\vecx,\yd);
\foreach \i/\txt in {0/{x_0^d},1/{\cdots},2/{x_{n-1}^d},3/{0},4/{\cdots},5/{0}}{
    \pgfmathsetmacro{\xx}{\i*\cellw}
    \draw[draw=blue!65!black,fill=blue!8]
        ([xshift=\xx cm]pdsw) rectangle ++(\cellw,\cellh);
    \node at ([xshift=\xx cm + 0.5*\cellw cm,yshift=0.5*\cellh cm]pdsw)
        {$\txt$};
}
\node[above=0.0mm] at ($(pdsw)+(3.0,0.50)$)
    {$\mathbf p_d=(x_0^d,\ldots,x_{n_1-1}^d,0,\ldots,0)$};

\foreach \p in {p0sw,p1sw,pdsw}{
    \draw[dashed,thin]
        ([xshift=3*\cellw cm]\p)
        -- ([xshift=3*\cellw cm,yshift=\cellh cm]\p);
}

\coordinate (p0west) at ($(p0sw)+(0,0.25)$);
\coordinate (p1west) at ($(p1sw)+(0,0.25)$);
\coordinate (pdwest) at ($(pdsw)+(0,0.25)$);
\coordinate (p0east) at ($(p0sw)+(6*\cellw,0.25)$);
\coordinate (p1east) at ($(p1sw)+(6*\cellw,0.25)$);
\coordinate (pdeast) at ($(pdsw)+(6*\cellw,0.25)$);

\node[coeff] (s0) at (-5.65,\yzero+0.25) {$\sigma_0$};
\node[coeff] (s1) at (-5.65,\yone+0.25) {$\sigma_1$};
\node[font=\Large] at (-5.65,-2.35) {$\vdots$};
\node[coeff] (sd) at (-5.65,\yd+0.25) {$\sigma_d$};

\node (m0) at (-4.42,\yzero+0.25) {$\otimes$};
\node (m1) at (-4.42,\yone+0.25) {$\otimes$};
\node (md) at (-4.42,\yd+0.25) {$\otimes$};

\node[font=\Large] at ($(p1sw)+(3.0,-0.2)$) {$\vdots$};

\draw[flow] (s0.east) -- (m0.west);
\draw[flow] (m0.east) -- (p0west);

\draw[flow] (s1.east) -- (m1.west);
\draw[flow] (m1.east) -- (p1west);

\draw[flow] (sd.east) -- (md.west);
\draw[flow] (md.east) -- (pdwest);

\node[acc] (accum) at (0,-4.45) {%
$\mathbf r=\sigma_0\mathbf p_0\oplus\sigma_1\mathbf p_1
\oplus\cdots\oplus\sigma_d\mathbf p_d$\\[0.5mm]
$=\bigl(\sigma(x_0),\ldots,\sigma(x_{n_1-1}),0,\ldots,0\bigr)$
};

\node[align=left] at (5.85,-4.45) {lane-wise XOR\\accumulation};

\coordinate (b0) at (\busx,\yzero+0.25);
\coordinate (b1) at (\busx,\yone+0.25);
\coordinate (bd) at (\busx,\yd+0.25);
\coordinate (ba) at (\busx,-4.45);

\draw[plain] (p0east) -- (b0);
\draw[plain] (p1east) -- (b1);
\draw[plain] (pdeast) -- (bd);
\draw[plain] (b0) -- (ba);
\draw[flow] (ba) -- (accum.east);

\node[comp] (cmp) at (0,-5.60) {%
compare $r_i$ with $0$\\[-0.2mm]
\texttt{VCMPEQ}$(\mathbf r,0)$
};
\draw[flow] (accum) -- (cmp);

\node[mask] (emask) at (0,-6.70) {%
$\mathbf e^{\mathrm{loc}}=
\bigl(\mathbf 1\{r_0=0\},\mathbf 1\{r_1=0\},
\ldots,\mathbf 1\{r_{n_1-1}=0\}\bigr)$
};
\draw[flow] (cmp) -- (emask);

\node[align=left] at (5.75,-6.45) {root indicator /\\error-location mask};

\end{tikzpicture}%
}
\caption{Vectorized shortened-support Chien search on packed support points, where \(d=\deg\sigma\).}
\label{fig:vectorized-shortened-support-chien}
\end{figure}
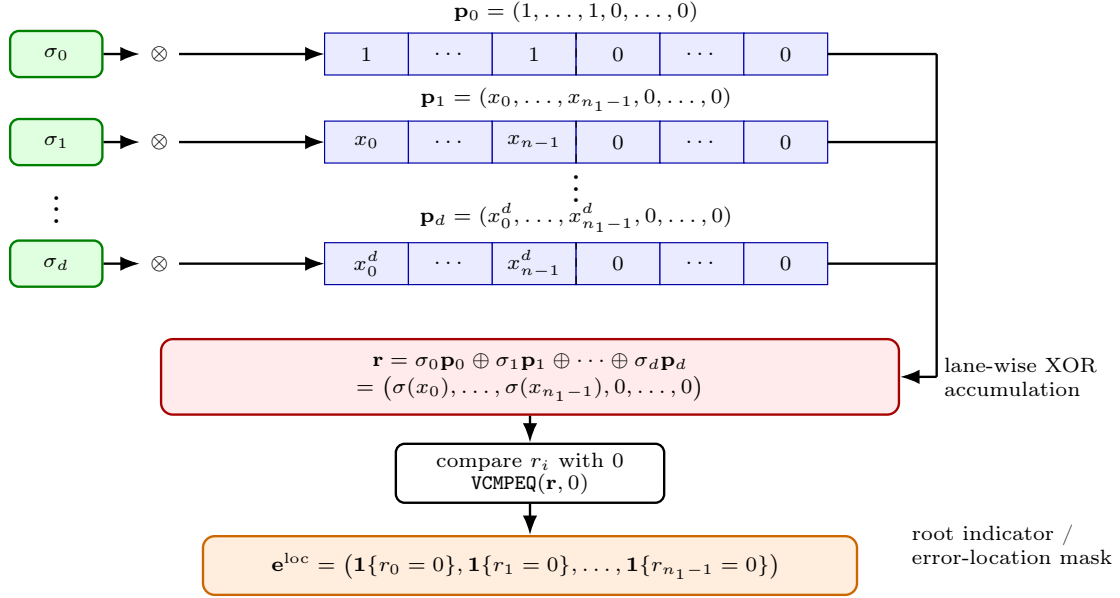

Thus, the loop over the locator coefficients remains sequential, but for each
coefficient all shortened support points are evaluated in parallel across HVX
lanes. Figure~\ref{fig:vectorized-shortened-support-chien} illustrates this procedure.

\subsubsection{Error-locator polynomial computation.}

The error-locator polynomial $\sigma(x)$ is recovered from the syndromes using the
Berlekamp--Massey algorithm~\cite{Berlekamp68}. Over $2\delta$ iterations, the algorithm
maintains the current locator $\sigma(x)$, an auxiliary (previous-best) locator
$b(x)$, and their degrees. At each iteration $\mu$ it (i)~computes a discrepancy as
the inner product of the current locator with the most recent syndromes,
\[
  d_\mu \;=\; S_\mu \;\oplus\; \bigoplus_{i=1}^{\deg \sigma} \sigma_i\, S_{\mu-i},
\]
and (ii)~when $d_\mu \neq 0$, updates the locator by
$\sigma(x) \leftarrow \sigma(x) \oplus \gamma\, x^{m} b(x)$, where
$\gamma = d_\mu \cdot d_\rho^{-1}$ and $d_\rho$ is the discrepancy recorded at the
last length change. Whenever the length-change condition $2\deg\sigma \le \mu$ holds,
the algorithm promotes the saved locator to the auxiliary polynomial $b(x)$, resets
the shift counter $m$, and updates the tracked degrees.

Unlike the Reed-Muller stages and the syndrome and root-search kernels, the
Berlekamp--Massey recurrence is inherently sequential: the discrepancy at iteration
$\mu$ depends on the locator coefficients produced by the preceding iterations, and
the locator and auxiliary polynomials are short---at most $\delta+1 = 16$ coefficients
for HQC-128. Vectorizing this recurrence across HVX lanes therefore yields little
benefit while complicating the control flow. We accordingly keep the error-locator
stage scalar and instead accelerate its dominant cost---finite-field
multiplication---using the table-driven logarithm/antilogarithm method of
Section~\ref{sec:gf-mul}. Each locator update and each discrepancy
term is a single $\mathbb{F}_{2^8}$ product reduced to two logarithm lookups, one
modular addition, and one antilogarithm lookup; the outer recurrence runs the full
$2\delta = 30$ iterations, while the inner discrepancy and update loops range only
over the active coefficients up to the current locator degree. With table-driven
arithmetic, this stage drops from $119{,}595$ to $6{,}581$ Pcycles per decode
(Table~\ref{tab:substage-pcycle-breakdown})---about $16\%$ of the optimized HQC-128 decode---without
any HVX vectorization. The full procedure is given in Algorithm~\ref{alg:elp}.


\subsection{Extension to HQC-192 and HQC-256}

\begin{table}[H]
\centering
\begin{tabular}{|c|c|c|c|}
\hline
Tweak                  & HQC-128 & HQC-192 & HQC-256 \\
\hline
Reed-Muller multiplicity        & 3       & 5       & 5       \\
\hline 
RS Code Length         & 46      & 56      & 90      \\
\hline 
Error Capacity         & 15      & 16      & 29      \\
\hline 
Chien HVX vectors      & 1       & 1       & 2       \\
\hline
\end{tabular}
\vspace{0.5em}
\caption{Extension to HQC-192 and HQC-256}
\label{tab:extension-to-HQC-192-256}
\end{table}

Although the implementation discussion above mainly focuses on the HQC-128 parameter set, the same optimization strategy extends naturally to HQC-192 and HQC-256. The three HQC variants differ primarily in the number of Reed–Muller repetitions, Reed–Solomon code dimensions, and the resulting decoding workload.

For HQC-192 and HQC-256, the Reed–Muller aggregation stage generalizes directly from multiplicity 3 to multiplicity 5. Instead of summing three duplicated Reed-Muller blocks before the Hadamard transform, the decoder accumulates five repeated blocks using the same lane-wise HVX addition operations. Since the Reed-Muller block size remains fixed at 128 bits, the Hadamard stages and peak-selection procedure remain unchanged; only the number of vector accumulation steps increases.

The Reed–Solomon decoding stage also scales naturally to the larger parameter sets. In particular, HQC-256 increases the shortened Reed–Solomon support size to 90, which exceeds the 64 halfword lanes of a single HVX vector. We therefore extend the shortened-support Chien search to a multi-vector accumulation scheme: the 90 support points are partitioned across two HVX vector accumulators, each holding 64 halfword lanes (with 38 padding lanes in the second). For each locator coefficient $\sigma_j$, the scalar-by-vector multiplication is performed independently on each accumulator and combined via lane-wise XOR; the final scan over the first 90 lanes returns the located-error mask. The same packed-vector accumulation strategy extends to two parallel
accumulators for HQC-256. The same multi-vector accumulation idea is also used when packed evaluations exceed one HVX vector. 


Finally, all finite-field lookup tables, alpha-power tables, shortened-support vectors, and generator-polynomial coefficients are precomputed at compile time (alpha tables, generator polynomials) or initialized once at first use (transposed power tables for the HVX paths). This allows the same decoding kernels to operate across HQC-128, HQC-192, and HQC-256 without modifying the underlying vectorized arithmetic routines.

\paragraph{Security scope.}
This work targets the performance and energy efficiency of HQC decoding on NPU-integrated Qualcomm platforms, and the evaluated backend is not constant-time. Two sources of secret-dependent behavior are present. First, following the table-driven Reed–Solomon optimization of OptHQC ~\cite{DongFW25OptHQC}, our finite-field arithmetic (Section \ref{sec:gf-mul}) replaces multiplications with logarithm/antilogarithm lookups, which use operand-dependent table indices. Second, the error-locator computation (Algorithm \ref{alg:elp}) is a branchy Berlekamp–Massey recurrence whose control flow and inner-loop bounds depend on the discrepancy values and the evolving locator degree. The HVX kernels for the Reed–Muller transform, peak selection, syndrome evaluation, and Chien search use fixed loop bounds and data-independent access patterns, but they operate on values derived from the (secret-dependent) received word. We therefore make no claim of resistance to address-dependent microarchitectural leakage or power/timing side channels. A fully constant-time backend—using fixed-flow finite-field arithmetic and a masked, branch-free error-locator computation—is left for future work.
 
\section{Experimental Evaluation}
\label{sec:experiment}

This section evaluates the performance and energy efficiency of the proposed NPU-supported HQC decoder. We first report simulator measurements for all three HQC parameter sets in terms of Hexagon processor cycles, and then provide a substage-level breakdown for HQC-128 to identify the main sources of improvement in the Reed-Muller and Reed-Solomon decoding components.
Finally, we present real-device measurements on a Snapdragon 8 Gen 2 development platform for all three HQC parameter sets, reporting latency, energy per decode, throughput per watt, and host CPU utilization. Our proof-of-concept implementation is available at
\href{https://github.com/Hiiamming/pqc-hqc-npu}{github.com/Hiiamming/pqc-hqc-npu}.

Across all simulator and real-device runs reported below, the optimized decoder produced output identical to the reference scalar decoder on the full 256-fixture corpus for every parameter set, confirming the bit-level equivalence targeted in Section \ref{sec:proposed}. All fixture corpora used for decoding are generated randomly at each iteration. We refer readers to our proof-of-concept implementation for further details.

\subsection{Experiment Settings}

\subsubsection{Environment}
We benchmark the implementation in two execution environments: the cycle-accurate \texttt{hexagon-sim} shipped with the Qualcomm Hexagon SDK, and
the on-device cDSP path accessed from an Android host through FastRPC. The simulator binary is executed through the H2 minimal hypervisor and its booter
entrypoint. The cDSP image is compiled with the same \texttt{hexagon-clang} configuration as the simulator binary, while the Android host benchmark is
cross-compiled with the NDK Bionic toolchain. Throughout the evaluation, we use \emph{NPU-supported} to denote the cDSP/Hexagon/HVX backend, in contrast to the scalar CPU or scalar Hexagon baselines.


\subsubsection{ FastRPC communication with the NPU}
In the execution context, the CPU requires a communication protocol to offload specific tasks to the NPU. Since the CPU and NPU operate in different execution domains, the CPU cannot directly invoke NPU kernels as ordinary local functions. Instead, it relies on a runtime communication mechanism, namely FastRPC~\cite{Qua18HexagonDSP}, to transfer control information, pass buffer descriptors, and synchronize task execution.

\begin{figure}[h]

\centering
\begin{tikzpicture}[
  cpuNode/.style={
    draw=blue!70!black, fill=blue!8, text=blue!60!black,
    rounded corners=6pt, minimum width=3.8cm, minimum height=1.1cm,
    align=center, font=\small\bfseries, line width=0.6pt
  },
  bufNode/.style={
    draw=cyan!70!black, fill=cyan!8, text=cyan!60!black,
    rounded corners=6pt, minimum width=3.8cm, minimum height=1.3cm,
    align=center, font=\small\bfseries, line width=0.6pt
  },
  stubNode/.style={
    draw=magenta!70!black, fill=magenta!8, text=magenta!60!black,
    rounded corners=6pt, minimum width=3.8cm, minimum height=1.3cm,
    align=center, font=\small\bfseries, line width=0.6pt
  },
  drvNode/.style={
    draw=yellow!50!black, fill=yellow!12, text=yellow!40!black,
    rounded corners=6pt, minimum width=3.8cm, minimum height=1.1cm,
    align=center, font=\small\bfseries, line width=0.6pt
  },
  dspNode/.style={
    draw=red!70!black, fill=red!8, text=red!60!black,
    rounded corners=6pt, minimum width=3.8cm, minimum height=1.1cm,
    align=center, font=\small\bfseries, line width=0.6pt
  },
  kernNode/.style={
    draw=green!60!black, fill=green!8, text=green!45!black,
    rounded corners=6pt, minimum width=3.8cm, minimum height=1.1cm,
    align=center, font=\small\bfseries, line width=0.6pt
  },
  solidArr/.style={-{Stealth[length=5pt,width=4pt]}, thick},
  dashArr/.style={-{Stealth[length=5pt,width=4pt]}, thick, dashed, gray!70},
  lbl/.style={font=\scriptsize, fill=white, inner sep=2pt}
]

\node[cpuNode]  (cpu)  at (0,    0  ) {CPU Application};

\node[bufNode]  (buf)  at (0,   -2.4)
  {Shared I/O Buffer\\
   {\footnotesize\mdseries\color{cyan!60!black}DMA-accessible, zero-copy}};

\node[stubNode] (stub) at (0,   -5.0)
  {FastRPC Stub\\
   {\footnotesize\mdseries\color{magenta!60!black}User-space proxy, marshals args}};

\draw[
  draw=gray!60,
  dashed,
  rounded corners=8pt,
  fill=gray!10
] (-6.75,-8.95) rectangle (6.75,-6.95);

\node[
  font=\scriptsize\itshape,
  text=gray!70
] at (-5.35,-7.15) {NPU subsystem};

\node[drvNode]  (drv)  at (-4.4, -8.0)
  {FastRPC Driver\\
   {\footnotesize\mdseries\color{yellow!40!black}Kernel driver, IPC}};

\node[dspNode]  (dsp)  at ( 0,   -8.0)
  {NPU Runtime\\
   {\footnotesize\mdseries\color{red!60!black}Schedules compute tasks}};

\node[kernNode] (kern) at ( 4.4, -8.0)
  {Kernel Execution\\
   {\footnotesize\mdseries\color{green!45!black}Runs NN / signal kernels}};

\draw[solidArr, blue!70!black]
  (cpu.south) -- (buf.north)
  node[lbl, midway, right=5pt] {write input};

\draw[solidArr, cyan!70!black]
  (buf.south) -- (stub.north)
  node[lbl, midway, right=5pt] {invoke with buffer};

\draw[solidArr, magenta!70!black]
  (stub.south) -- ++(0,-0.6) -| (drv.north);

\draw[solidArr, yellow!50!black]
  (drv.east) -- (dsp.west);

\draw[solidArr, red!70!black]
  (dsp.east) -- (kern.west);

\coordinate (A) at (buf.east);
\coordinate (B) at (6.9, -2.4);
\coordinate (C) at (6.9, -8.0);
\coordinate (D) at (kern.east);

\draw[dashArr]
  (A) -- (B) -- (C) -- (D);

\node[lbl, rotate=90] at ($(B)!0.5!(C)$)
  {buffer handle / descriptor};

\coordinate (E) at (kern.south);
\coordinate (F) at (4.4,  -9.4);
\coordinate (G) at (-7.0, -9.4);
\coordinate (H) at (-7.0, -5.0);
\coordinate (I) at (stub.west);

\draw[dashArr]
  (E) -- (F) -- (G) -- (H) -- (I);

\node[lbl] at ($(F)!0.5!(G)+(0,-0.05)$) [below=2pt]
  {completion / status};

\end{tikzpicture}
\caption{FastRPC-based NPU task execution flow.}
\label{fig:fastrpc_base_NPU}
\end{figure}
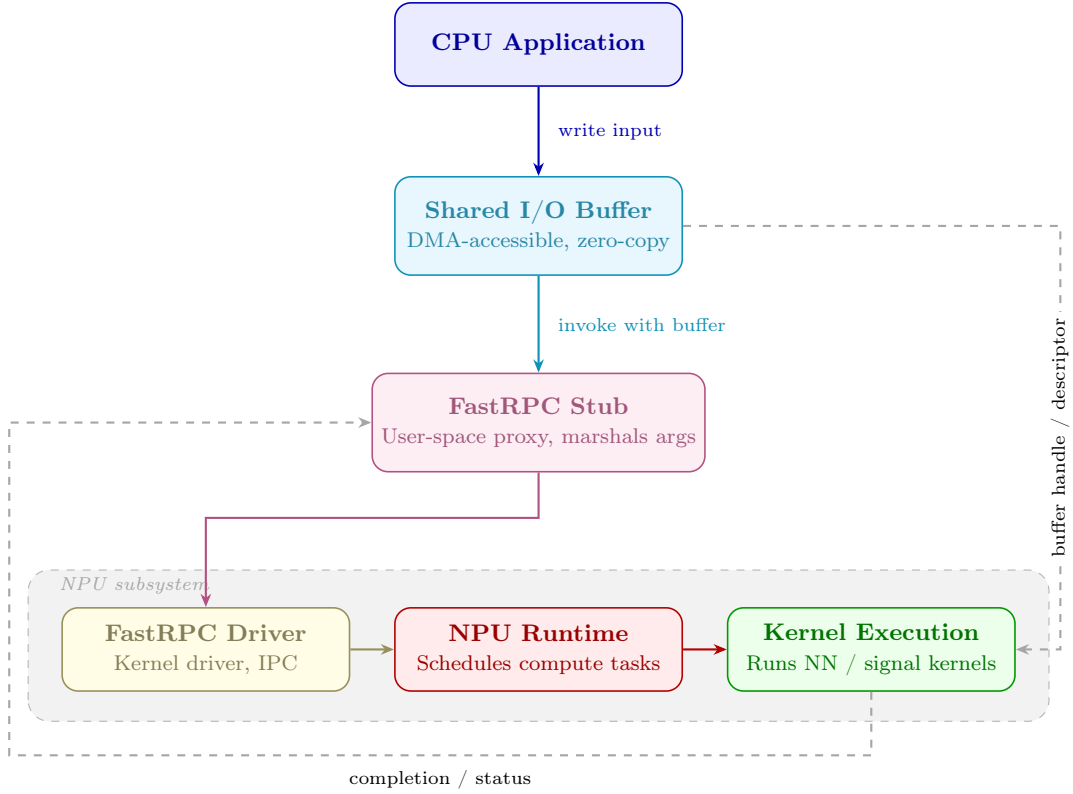

In this workflow, the CPU first prepares the input data in a shared I/O buffer that is accessible by both the CPU and the NPU. Rather than copying large data through the control path, the CPU passes only lightweight metadata, such as buffer handles, data sizes, and task parameters. The FastRPC stub on the CPU side marshals these arguments and forwards the request to the NPU-side runtime. The NPU runtime then dispatches the corresponding kernel, reads the input from the shared buffer, performs the computation, and writes the result back to the output buffer. Finally, the completion status is returned to the CPU, which synchronizes and reads the output. The complete workflow is visualized in Figure~\ref{fig:fastrpc_base_NPU}.

Unfortunately, each initialization incurs a non-negligible overhead, averaging around $557~\mu$s, which makes per-decode initialization undesirable. To address this issue, we batch multiple decoding instances and offload them to the NPU in a single communication round, thereby amortizing the extra cost. This strategy is natural in practical applications, where devices often process continuously arriving data, especially in streaming video workloads. The detailed measurement is given in Table~\ref{tab:fastrpc-boundary-overhead}.

\subsubsection{Measurement metrics}
We report four complementary metrics to evaluate the proposed implementation: simulated processor cycles, real-device latency, device energy, and CPU
utilization.

\paragraph{Hexagon Pcycles.}
To avoid simulator startup and teardown effects, we use a paired measurement strategy. We measure both a one-iteration and a three-iteration run, subtract the former from the latter, and normalize by the number of additional decodes:
\[
    \mathrm{Pcycles/decode}
    =
    \frac{
        \mathrm{Pcycles}_{T=3}
        -
        \mathrm{Pcycles}_{T=1}
    }{
        (3-1) \times N_{\mathrm{fix}}
    } .
\]
where $N_{\mathrm{fix}} = 256$ is the corpus size. The same estimator is used for both the scalar baseline and the NPU-supported
variant, as well as for the substage breakdown.

\paragraph{Real-device latency and energy.}
On the Android device, we measure wall-clock latency around the decoding loop and divide by the total number of decodes. Energy is measured by sampling device power-supply voltage and current during execution, integrating power over time, and normalizing by the number of decodes. We disable \texttt{qprof} during energy measurement to avoid perturbing power and clock behavior.

\paragraph{CPU utilization.}
We measure the CPU cost paid by the Android host process under each backend. In the CPU-scalar configuration, this is the process that runs the scalar decoder directly on the Kryo CPU. In the NPU-supported configuration, it is the same host-side benchmark process acting as a FastRPC dispatcher: it submits batched decode work to the cDSP and waits for completion. The measurement wrapper samples \texttt{/proc/<pid>/task/*/stat} while the process is alive, sums the user and system jiffies across all of the process threads, and converts the tick delta using \texttt{getconf CLK\_TCK}. We then normalize the resulting host CPU time in two ways. The process CPU percentage divides host CPU time by the wall-clock interval, so \(100\%\) corresponds to one fully loaded CPU core on the Snapdragon 8 Gen 2 platform rather than to the entire multi-core SoC. The CPU ms/decode metric divides the same host CPU time by the number of completed decodes. Finally, CPU reduction compares CPU ms/decode against the CPU-scalar baseline and therefore quantifies host-side offload only; NPU-supported decoding still consumes Hexagon cycles on the cDSP, which are reflected separately in the latency and energy measurements.

\subsection{Simulation Results}

The simulator benchmark is conducted on the current 256-fixture decoding corpus
for each HQC parameter set and reports Hexagon processor cycles (Pcycles).
Instead of reporting raw \texttt{hexagon-sim} totals as the main result, we use
the paired \(T=1\) and \(T=3\) runs to expose the actual normalized cost. The
\(\Delta\) columns in Table~\ref{tab:sim-pcycles-full} are the difference between
the three-iteration and one-iteration runs; the estimated cycles per decode are
then obtained by dividing \(\Delta\)Pcycles by \(\Delta\)decodes. This presentation
keeps the fixed simulator/setup overhead separate from the per-decode estimate.

\begin{table}[htbp]
\centering
\renewcommand{\arraystretch}{1.15}
\small
\begin{tabular}{|l|l|c|c|c|c|}
\hline
\textbf{Parameter} & \textbf{Backend} & \(\Delta\)\textbf{Pcycles} &
\(\Delta\)\textbf{decodes} & \textbf{Estimated Pcycles/decode} &
\textbf{Speedup} \\
\hline
HQC-128 & Scalar & 488,326,812 & 512 & 953,763 & 1.00x \\
\hline
HQC-128 & NPU-supported & 21,233,214 & 512 & 41,471 & 23.00x \\
\hline
HQC-192 & Scalar & 678,800,028 & 512 & 1,325,781 & 1.00x \\
\hline
HQC-192 & NPU-supported & 25,714,332 & 512 & 50,223 & 26.40x \\
\hline
HQC-256 & Scalar & 1,405,108,650 & 512 & 2,744,353 & 1.00x \\
\hline
HQC-256 & NPU-supported & 41,295,582 & 512 & 80,655 & 34.03x \\
\hline
\end{tabular}
\vspace{0.5em}
\caption{Paired Hexagon simulator measurements for full HQC decoding. The
reported per-decode cost is \(\Delta\)Pcycles/\(\Delta\)decodes from the
\(T=1\) and \(T=3\) runs, not a raw total-cycle division.}
\label{tab:sim-pcycles-full}
\end{table}

Relative to the scalar baseline, the NPU-supported path is \(23.00\times\),
\(26.40\times\), and \(34.03\times\) faster for HQC-128, HQC-192, and HQC-256,
respectively.

\subsection{Substage Analysis}

The substage benchmark compares the scalar baseline with the NPU-supported path.
Reed-Muller substage costs are first estimated from paired one- and
three-iteration measurements per RM block, then multiplied by the \(46\)
Reed-Muller blocks used in one HQC-128 decode. Reed-Solomon substage costs
use the same paired estimate directly per decode. The table reports the main
optimized substages rather than an exhaustive decomposition of the full decoder.

\begin{table}[htbp]
\centering
\renewcommand{\arraystretch}{1.15}
\begin{tabular}{|l|c|c|}
\hline
\textbf{Substage} & \textbf{Scalar} & \textbf{NPU-supported} \\
\hline
Reed-Muller Hadamard & 263,175 & 17,950 \\
\hline
Reed-Muller find peak & 71,217 & 6,081 \\
\hline
Reed-Solomon syndrome & 162,312 & 3,517 \\
\hline
Reed-Solomon error-locator polynomial & 119,595 & 6,581 \\
\hline
\textbf{Full decode estimate} & \textbf{953,763} & \textbf{41,471} \\
\hline
\end{tabular}
\vspace{0.5em}
\caption{Paired Pcycle breakdown of selected optimized HQC-128 decoding substages. Reed-Muller rows are converted from per-block cost to per-decode cost by multiplying by 46 blocks. Note that this table does not include repetition aggregation, codeword loading, and other small substages.}
\label{tab:substage-pcycle-breakdown} 
\end{table}

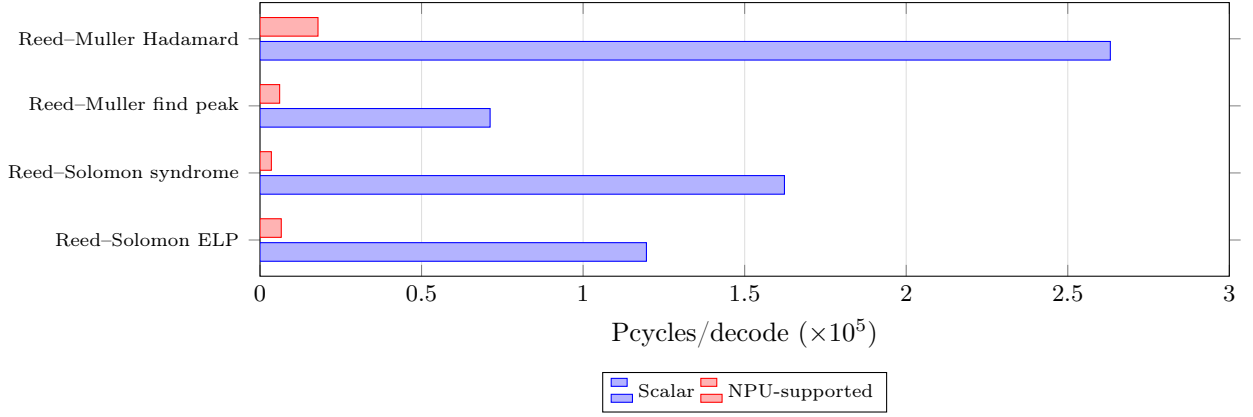
\begin{figure}[htbp] 
\centering
\begin{tikzpicture}
\begin{axis}[
    xbar,
    width=0.9\linewidth,
    height=5.2cm,
    bar width=7pt,
    xlabel={Pcycles/decode $(\times 10^5)$},
    xmin=0,
    xmax=3.0,
    xtick={0,0.5,1.0,1.5,2.0,2.5,3.0},
    scaled x ticks=false,
    symbolic y coords={
        Reed--Muller Hadamard,
        Reed--Muller find peak,
        Reed--Solomon syndrome,
        Reed--Solomon ELP
    },
    ytick=data,
    y dir=reverse,
    enlarge y limits=0.18,
    legend style={
        at={(0.5,-0.35)},
        anchor=north,
        legend columns=2,
        font=\scriptsize
    },
    x tick label style={font=\small},
    yticklabel style={font=\scriptsize},
    xmajorgrids=true,
    grid style={gray!25},
]
\addplot coordinates {
    (2.63175,Reed--Muller Hadamard)
    (0.71217,Reed--Muller find peak)
    (1.62312,Reed--Solomon syndrome)
    (1.19595,Reed--Solomon ELP)
};
\addplot coordinates {
    (0.17950,Reed--Muller Hadamard)
    (0.06081,Reed--Muller find peak)
    (0.03517,Reed--Solomon syndrome)
    (0.06581,Reed--Solomon ELP)
};
\legend{Scalar,NPU-supported}
\end{axis}
\end{tikzpicture}
\vspace{0.5em}
\caption{Comparison of selected optimized decoding substages between the scalar baseline and the NPU-supported backend (ELP is abbreviated for error locator polynomial). Lower is better.}
\label{fig:substage-pcycle-comparison}
\end{figure}

\subsection{Real-Device Results}

The real-device experiments were conducted on a Snapdragon\textregistered{} 8 Gen 2 Mobile Hardware Development Kit, model HDK8550, based on the Qualcomm
Snapdragon SM8550P application processor. According to the manufacturer specification, this platform includes an 8-core 64-bit Kryo CPU, consisting of
one Arm Cortex-X3 prime core up to 3.2 GHz, four performance cores up to 2.8 GHz, and three efficiency cores up to 2.0 GHz. It also includes a Qualcomm
Hexagon processor equipped with Hexagon Vector eXtensions (HVX), scalar and tensor accelerators, micro tile inferencing, Hexagon Direct Link, and support
for INT4, INT8, INT16, and FP16 arithmetic. Note that the simulator speedup is measured against a Hexagon scalar baseline, whereas the real-device speedup is against the ARM Cortex CPU scalar baseline; the two are not directly comparable.

The real-device measurements use the corresponding 256-fixture decoding corpus for each HQC parameter set. Unlike the simulator benchmark, which reports
Hexagon Pcycles, the real-device benchmark reports host-observed latency and direct device energy. Each reported value is the mean of five runs, and
each run performs \(32{,}000\) decodes. The NPU-supported backend uses the current non-worker FastRPC path, where one batched remote call executes the
full decode loop on the cDSP. Energy is measured from Android power-supply voltage/current samples with qprof disabled. Here, qprof refers to Qualcomm's profiler for DSP/NPU
utilization and clock-state diagnostics; its measurements are used only as diagnostic evidence and are not used for the energy claims.

\begin{table}[htbp]
\centering
\renewcommand{\arraystretch}{1.15}

\begin{tablenotes}
\footnotesize
\item The speedup, decodes/s/W, and energy-gain columns are computed per run
and then averaged over the five runs. Therefore, the reported speedup and
energy-gain values are means of per-run ratios, not ratios recomputed from the
rounded mean latency or mean energy columns. In particular, the energy-gain
column can differ from
\(\overline{E}_{\mathrm{CPU}} / \overline{E}_{\mathrm{NPU}}\)
because direct-energy measurements have higher run-to-run variability.
\end{tablenotes}

\begin{tabular}{|c|c|c|c|c|c|c|}
\hline
\textbf{HQC} & \textbf{Backend} & \textbf{us/decode} & \textbf{uJ/decode} &
\textbf{decodes/s/W} & \textbf{Speedup} & \textbf{Energy gain} \\
\hline
HQC-128 & CPU scalar    & 81.144  & 189.108 & 5,641.362  & 1.00x & 1.0x  \\
\hline
HQC-128 & NPU-supported & 39.173  & 10.593  & 110,083.297 & 2.07x & 18.13x \\
\hline
HQC-192 & CPU scalar    & 103.531 & 246.480 & 4,258.922  & 1.00x & 1.0x  \\
\hline
HQC-192 & NPU-supported & 56.065  & 23.645  & 53,040.740 & 1.85x & 11.77x \\
\hline
HQC-256 & CPU scalar    & 228.082 & 584.573 & 1,747.860  & 1.00x & 1.0x  \\
\hline
HQC-256 & NPU-supported & 116.333 & 36.078  & 30,497.824 & 1.96x & 16.81x \\
\hline
\end{tabular}
\vspace{0.5em}

\caption{Real-device latency, energy, and throughput-per-watt measurements across HQC parameter sets. Values are means over five runs; run-to-run variability is summarized separately in Table~\ref{tab:real-device-stability}.}
\label{tab:real-device-latency-energy}
\end{table}

The decodes/s/W column is the per-run throughput (decodes/s, from the reported per-decode latency) divided by the net active power (run minus idle power); like the energy-gain column, it is computed per run and then averaged over the five runs. Across all three HQC parameter sets, the NPU-supported backend improves both latency and energy efficiency over the scalar CPU baseline. The measured speedup is \(1.85\times\) to \(2.07\times\), and the energy gain is \(11.77\times\) to \(18.13\times\). On the current device run, NPU-supported latency increases with the HQC parameter set size, while still reducing both wall-clock time and energy per decode relative to the CPU baseline.

\begin{table}[htbp]
\centering
\renewcommand{\arraystretch}{1.15}
\small
\begin{tabular}{|c|c|c|c|c|}
\hline
\textbf{HQC} & \textbf{Backend} & \textbf{us/decode CV} &
\textbf{uJ/decode CV} & \textbf{CPU ms/decode CV} \\
\hline
HQC-128 & CPU scalar    & 0.17\% & 3.72\% & 0.83\% \\
\hline
HQC-128 & NPU-supported & 0.05\% & 14.84\% & 59.27\% \\
\hline
HQC-192 & CPU scalar    & 0.08\% & 4.59\% & 0.85\% \\
\hline
HQC-192 & NPU-supported & 0.21\% & 34.40\% & 81.44\% \\
\hline
HQC-256 & CPU scalar    & 0.10\% & 3.33\% & 0.32\% \\
\hline
HQC-256 & NPU-supported & 0.07\% & 18.66\% & 22.83\% \\
\hline
\end{tabular}
\vspace{0.5em}
\caption{Run-to-run variability for the direct real-device measurements, reported as coefficient of variation \(\mathrm{CV}=\sigma/\mu\). Latency is stable across runs; energy has higher variability because it is derived from board-level power samples.}
\label{tab:real-device-stability}
\end{table}

\begin{figure}[H]
\centering
\begin{tikzpicture}
\begin{axis}[
    ybar,
    width=\linewidth,
    height=8cm,
    bar width=14pt,
    ylabel={uJ/decode},
    symbolic x coords={HQC-128,HQC-192,HQC-256},
    xtick=data,
    ymin=0,
    ymax=800,
    enlarge x limits=0.18,
    legend style={at={(0.5,-0.15)},anchor=north,legend columns=2,font=\scriptsize},
    nodes near coords,
    nodes near coords style={font=\scriptsize,rotate=90,anchor=west},
    yticklabel style={/pgf/number format/fixed},
    x tick label style={font=\small},
]
\addplot coordinates {
    (HQC-128,189.108)
    (HQC-192,246.480)
    (HQC-256,584.573)
};

\addplot coordinates {
    (HQC-128,10.593)
    (HQC-192,23.645)
    (HQC-256,36.078)
};

\legend{CPU scalar,NPU-supported}
\end{axis}
\end{tikzpicture}
\caption{Real-device energy per decode. Lower is better.}
\label{fig:real-device-energy}
\end{figure}
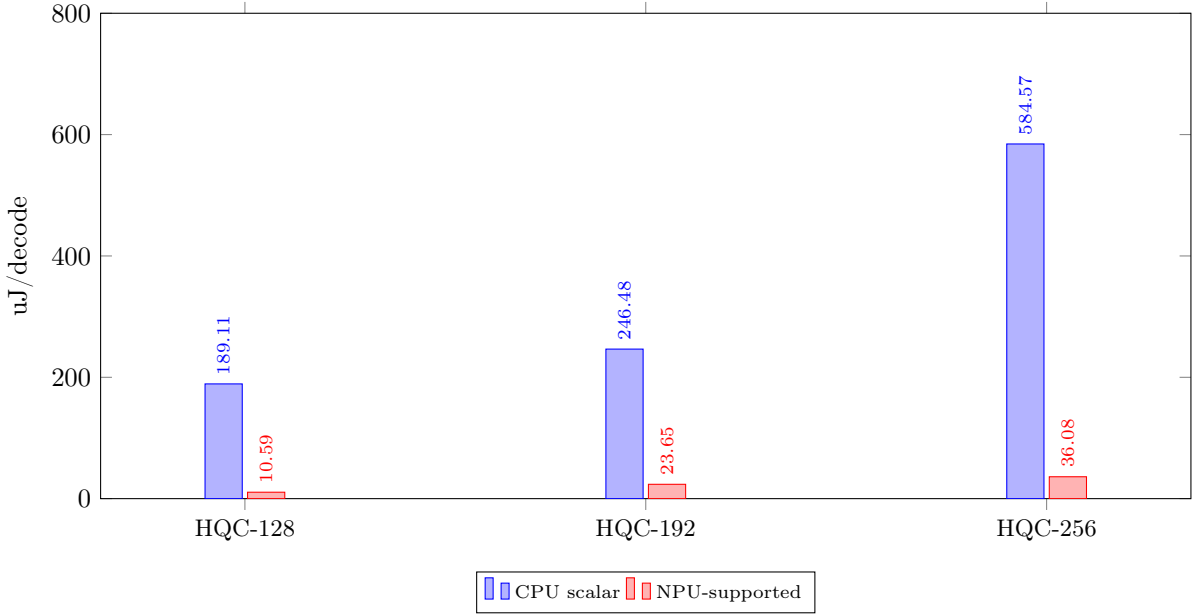

In addition to latency and energy, Table~\ref{tab:real-device-cpu-offload}
reports the CPU time consumed by the benchmark host process itself. These
numbers are collected by wrapping the benchmark with
\texttt{measure\_process\_cpu.sh}. The script samples the process and
thread accounting files under \texttt{/proc/<pid>/task/*/stat} while the
benchmark is running, sums user and system CPU ticks, and converts them to
milliseconds using the Linux clock tick rate. The same wrapper is used for both
backends: for CPU scalar, the measured process runs the scalar decoder directly;
for NPU-supported, the measured process mostly submits FastRPC work to the cDSP
and waits for completion.

The \textbf{process CPU \%} column is the sampled host CPU time divided by the
wall-clock runtime of the benchmark. It is normalized to one CPU core: \(100\%\)
means the process kept one CPU core fully busy on average, not the whole
multi-core SoC. The \textbf{CPU ms/decode} column is the same host CPU time
divided by the number of completed decodes. The \textbf{CPU reduction} column is
computed from CPU ms/decode as
\[
100 \times
\left(
1 -
\frac{\mathrm{CPU\ ms/decode}_{\mathrm{NPU}}}
     {\mathrm{CPU\ ms/decode}_{\mathrm{scalar}}}
\right).
\]

The reported CPU-reduction values are computed with this formula for each paired run and then averaged. They should therefore be interpreted as the mean
per-run host-side CPU offload, not as a value recomputed from the rounded mean
CPU ms/decode columns in the table

This metric measures host-side CPU offload only; the NPU-supported decoder
still consumes Hexagon cycles on the cDSP.

\begin{table}[htbp]
\centering
\renewcommand{\arraystretch}{1.15}
\small
\begin{tabular}{|c|c|c|c|c|}
\hline
\textbf{HQC} & \textbf{Backend} & \textbf{process CPU \%} &
\textbf{CPU ms/decode} & \textbf{CPU reduction} \\
\hline
HQC-128 & CPU scalar    & 93.606 & 0.0803125 & 0.000\% \\
\hline
HQC-128 & NPU-supported & 1.498  & 0.0006875 & 99.145\% \\
\hline
HQC-192 & CPU scalar    & 94.538 & 0.1025625 & 0.000\% \\
\hline
HQC-192 & NPU-supported & 0.678  & 0.0004375 & 99.575\% \\
\hline
HQC-256 & CPU scalar    & 97.243 & 0.2268750 & 0.000\% \\
\hline
HQC-256 & NPU-supported & 0.606  & 0.0007500 & 99.669\% \\
\hline
\end{tabular}
\vspace{0.5em}
\caption{Process-level CPU utilization and CPU-time reduction on the real device.}
\label{tab:real-device-cpu-offload}

\end{table}

The process-level CPU measurements show why CPU offloading matters beyond energy
alone. When the CPU consumes nearly all of one core decoding HQC, the host has
less room for other application or system work. Moving the decoding workload to
the NPU reduces host CPU ms/decode by \(99.1\%\)--\(99.7\%\) across all HQC
parameter sets. This reduction is a host-side offload metric only; the
NPU-supported decode still consumes Hexagon cycles, which are reflected in the
latency and energy measurements above.

We also isolate the FastRPC boundary cost. The batched row uses the same
non-worker batched decode benchmark as the real-device latency table, while the
``decode-one'' experiment makes one FastRPC call per decode. Table
\ref{tab:fastrpc-boundary-overhead} reports only the boundary-sensitive modes:
a no-op ping, one-decode-per-RPC, and the batched path used to
quantify boundary amortization.

\begin{table}[htbp]
\centering
\renewcommand{\arraystretch}{1.15}
\small

\begin{tabular}{|c|c|c|c|c|}
\hline
\textbf{HQC} & \textbf{Ping us/RPC} & \textbf{Decode-one us/decode} &
\textbf{Batched us/decode} & \textbf{Decode-one slowdown} \\
\hline
HQC-128 & 552.804 & 590.640 & 39.205 & 15.07x \\
\hline
HQC-192 & 530.990 & 604.201 & 56.234 & 10.74x \\
\hline
HQC-256 & 535.708 & 629.596 & 116.104 & 5.42x \\
\hline
\end{tabular}
\vspace{0.5em}
\caption{FastRPC boundary overhead on the real device. Values are means over five runs; variability is summarized separately in Table~\ref{tab:fastrpc-boundary-stability}.}
\label{tab:fastrpc-boundary-overhead}
\end{table}

\begin{table}[htbp]
\centering
\renewcommand{\arraystretch}{1.15}
\small
\begin{tabular}{|c|c|c|c|}
\hline
\textbf{HQC} & \textbf{Ping CV} & \textbf{Decode-one CV} &
\textbf{Batched CV} \\
\hline
HQC-128 & 1.47\% & 1.93\% & 0.08\% \\
\hline
HQC-192 & 2.71\% & 1.03\% & 0.68\% \\
\hline
HQC-256 & 2.73\% & 2.10\% & 0.11\% \\
\hline
\end{tabular}
\vspace{0.5em}
\caption{Run-to-run variability for the FastRPC boundary measurements, reported as coefficient of variation \(\mathrm{CV}=\sigma/\mu\).}
\label{tab:fastrpc-boundary-stability}
\end{table}

The FastRPC no-op boundary alone is about \(531\)--\(553\ \mu\mathrm{s}\) per call, which
is much larger than the batched per-decode cost for all three HQC parameter
sets. Calling FastRPC once per decapsulation is therefore \(5.42\times\) to
\(15.07\times\) slower than batching the decode loop inside one remote call.
Representative cached payload movement remains secondary to the boundary
itself: the measured cached input overhead is on the order of \(575\)--\(594\ \mu\mathrm{s}\)
per RPC for the current build. Thus, the real-device NPU numbers
should be interpreted as batched cDSP throughput with heavily amortized FastRPC
overhead, not as the latency of an API that performs one remote call per
decapsulation.

\section{Conclusion}
\label{sec:conclusion}
This paper presented an optimized implementation of HQC decoding on Qualcomm Hexagon processors in NPU-integrated devices. We showed that HQC decoding is well matched to the Hexagon/HVX vector execution model, since its dominant Reed-Muller and Reed-Solomon components naturally operate on vector-structured data such as reliability vectors, Hadamard-transform coefficients, syndrome vectors, finite-field elements, and packed support points. Based on this observation, we redesigned the main decoding kernels around HVX-friendly data layouts and execution patterns, including a vectorized Reed-Muller Hadamard transform, scalar-equivalent peak selection, HVX-oriented finite-field multiplication vectorized syndrome computation, and shortened-support Chien search.

Our evaluation demonstrates that the proposed Hexagon/HVX-assisted implementation substantially reduces the cost of HQC decoding. In Hexagon simulator measurements for HQC-128, the optimized implementation reduces the full decoding cost from $953{,}763$ to $41{,}471$ Pcycles per decode, corresponding to a $23.00\times$ speedup and a $95.7\%$ reduction in Pcycles; the same approach yields $26.40\times$ and $34.03\times$ speedups for HQC-192 and HQC-256, respectively. On the real device, where the NPU path additionally pays the FastRPC boundary cost, the batched backend improves end-to-end latency over the ARM CPU scalar baseline by $2.07\times$, $1.85\times$, and $1.96\times$, and improves direct-energy efficiency by $18.13\times$, $11.77\times$, and $16.81\times$ for HQC-128, HQC-192, and HQC-256, respectively. At the same time, it reduces host-side CPU time per decode by $99.1\%$--$99.7\%$, freeing the application processor for other work while the decode runs on the cDSP. Because a single FastRPC boundary crossing ($531$--$553~\mu s$) dominates one isolated decode, and one-decode-per-call execution costs about $591$--$630~\mu s$ per decode, these real-device gains are obtained by batching the decode loop into one remote call; the reported numbers should therefore be read as batched cDSP throughput with amortized FastRPC overhead rather than as the latency of a one-decode-per-call interface.

These results indicate that NPU-integrated mobile platforms can provide an effective execution backend not only for AI-oriented workloads, but also for structured post-quantum cryptographic decoding when the underlying kernels are reformulated around the available vector execution model. Our optimizations target the concatenated decoding stage of HQC; combining them with an accelerated sparse polynomial product for the full decapsulation path is a natural next step on the same backend. As future work, we plan to investigate fully constant-time finite-field arithmetic for all scalar and vector operations, strengthen the side-channel resilience of the backend, integrate a worker-pool cDSP schedule to further amortize host coordination, and extend the same hardware-aware optimization methodology to other code-based and post-quantum cryptographic primitives.

\paragraph{Declaration of Generative AI and AI-assisted technologies in the writing process.} During the preparation of this work, the authors used AI-assisted tools to assist with language editing, manuscript polishing, and code drafting/debugging for the experimental implementation. After using these tools, the authors reviewed, edited, and verified the manuscript content, implementation, and experimental results as necessary, and take full responsibility for the content of the publication.

\clearpage

\begingroup
\sloppy
\bibliographystyle{unsrt} 

\bibliography{crypto}
\endgroup

\clearpage

\section*{Appendix}
\appendix
\section{Supplementary Implementation Details}
\label{sec:appendix}
This section provides supplementary implementation details for the optimized HQC decoder. We first present the pseudocode of the main HVX-assisted routines used in the Reed-Muller and Reed-Solomon decoding stages. We then summarize the mapping between the abstract vector operations used throughout the paper and representative Qualcomm Hexagon HVX intrinsics.
\subsection{Pseudocode}

\begin{algorithm}[H]
\caption{HVX-based Hadamard transform for Reed-Muller decoding.}
\label{alg:hadamard-hvx}
\begin{algorithmic}[1]
\Statex \textbf{Input:} Source buffer \(\mathsf{src}\in\mathbb Z^{128}\) and temporary buffer \(\mathsf{dst}\in\mathbb Z^{128}\) of signed 16-bit values.
\Statex \textbf{Output:} Hadamard-transformed vector stored in the buffer pointed to by \(p_1\).

\State \(p_1 \gets \mathsf{src},\quad p_2 \gets \mathsf{dst}\)
\For{\(\mathsf{pass}=0\) \textbf{to} \(6\)}
    \State \(\mathbf{lo} \gets p_1[0:63],\quad \mathbf{hi}\gets p_1[64:127]\)
    \State \((\mathbf d_{lo},\mathbf d_{hi})
        \gets \mathsf{VDEAL2}(\mathbf{hi},\mathbf{lo},2)\)
        \Comment{two-vector deinterleaving}
    \State \(\mathbf v_e \gets \mathsf{VDEALH}(\mathbf d_{lo})\)
        \Comment{pack even lanes}
    \State \(\mathbf v_o \gets \mathsf{VDEALH}(\mathbf d_{hi})\)
        \Comment{pack odd lanes}
    \State \(\mathbf{sum}\gets \mathsf{VADDH}(\mathbf v_e,\mathbf v_o)\)
        \Comment{sum half}
    \State \(\mathbf{diff}\gets \mathsf{VSUBH}(\mathbf v_e,\mathbf v_o)\)
        \Comment{difference half}
    \State \(p_2[0:63]\gets \mathbf{sum},\quad p_2[64:127]\gets \mathbf{diff}\)
    \State \(\mathsf{swap}(p_1,p_2)\)
        \Comment{next pass consumes the output}
\EndFor
\State \Return \(p_1\)
\end{algorithmic}
\end{algorithm}

\begin{algorithm}[H]
\caption{HVX-based peak detection. }
\label{alg:find-peaks-hvx}
\begin{algorithmic}[1]
\Statex \textbf{Input:} Hadamard-transformed buffer
\(\mathsf{transform}\in\mathbb Z^{128}\) of signed 16-bit values.
\Statex \textbf{Output:} Encoded peak position for Reed--Muller decoding.

\State \(\mathbf{row}_0 \gets \mathsf{transform}[0:63]\),
       \(\mathbf{row}_1 \gets \mathsf{transform}[64:127]\)
       \Comment{load two HVX vectors}
\State \(\mathbf{abs}_0 \gets \mathsf{VABS}(\mathbf{row}_0)\),
       \(\mathbf{abs}_1 \gets \mathsf{VABS}(\mathbf{row}_1)\)
       \Comment{lane-wise magnitudes}
\State \(\mathbf{max} \gets \mathsf{VMAX}(\mathbf{abs}_0,\mathbf{abs}_1)\)
       \Comment{pairwise maximum}

\For{\(s\in\{64,32,16,8,4,2\}\)}
    \State \(\mathbf{max}\gets
    \mathsf{VMAX}(\mathbf{max},\mathsf{VROT}_{s}(\mathbf{max}))\)
    \Comment{rotation-based max reduction}
\EndFor

\State \(a^\star \gets \mathbf{max}[0]\)
       \Comment{global maximum magnitude}
\State \(\mathbf{target}\gets \mathsf{VSPLAT}(a^\star)\)
\State \(\mathbf{sentinel}\gets \mathsf{VSPLAT}(0x7fff)\)
       \Comment{invalid large index}
\State \(\mathbf{idx}_0\gets \mathsf{rm\_index\_lo}\),
       \(\mathbf{idx}_1\gets \mathsf{rm\_index\_hi}\)

\State \(\mathbf{mask}_0\gets \mathsf{VCMPEQ}(\mathbf{abs}_0,\mathbf{target})\)
\State \(\mathbf{mask}_1\gets \mathsf{VCMPEQ}(\mathbf{abs}_1,\mathbf{target})\)
       \Comment{mark tied maxima}

\State \(\mathbf{pos}_0\gets
\mathsf{VSELECT}(\mathbf{mask}_0,\mathbf{idx}_0,\mathbf{sentinel})\)
\State \(\mathbf{pos}_1\gets
\mathsf{VSELECT}(\mathbf{mask}_1,\mathbf{idx}_1,\mathbf{sentinel})\)
       \Comment{keep only valid peak indices}

\State \(\mathbf{peak}\gets \mathsf{VMIN}(\mathbf{pos}_0,\mathbf{pos}_1)\)

\For{\(s\in\{64,32,16,8,4,2\}\)}
    \State \(\mathbf{peak}\gets
    \mathsf{VMIN}(\mathbf{peak},\mathsf{VROT}_{s}(\mathbf{peak}))\)
    \Comment{minimum-index tie break}
\EndFor

\State \(i^\star\gets \mathbf{peak}[0]\)
\State \(v^\star\gets \mathsf{transform}[i^\star]\)
\If{\(v^\star>0\)}
    \State \(i^\star\gets i^\star\;|\;128\)
    \Comment{encode sign bit}
\EndIf
\State \Return \(i^\star\)
\end{algorithmic}
\end{algorithm}

\begin{algorithm}[H]
\caption{($\mathsf{XTIME\_VEC}$) Vectorized multiplication by $x$ over $\mathbb{F}_{2^8}$}
\label{alg:gf-xtime-hvx}
\begin{algorithmic}[1]
\Require Vector $\mathbf{x}$ whose lanes contain elements of $\mathbb{F}_{2^8}$
\Ensure Vector $\mathbf{y}$ where $y_i=X\cdot x_i \bmod P(X)$, where $P(X)=X^8+X^4+X^3+X^2+1$.

\State $\mathbf{carry} \gets
    \mathsf{VAND}(\mathsf{VSHR}_{7}(\mathbf{x}),\mathsf{VSPLAT}(1))$
    \Comment{MSB of each lane}

\State $\mathbf{mask} \gets
    \mathsf{VSUB}(\mathsf{VZERO}(),\mathbf{carry})$
    \Comment{expand carry bits to masks}

\State $\mathbf{shifted} \gets \mathsf{VSHL}_{1}(\mathbf{x})$
    \Comment{multiply by $X$}

\State $\mathbf{red} \gets
    \mathsf{VAND}(\mathbf{mask},\mathsf{VSPLAT}(0x1d))$
    \Comment{select reduction term}

\State $\mathbf{y} \gets \mathsf{VXOR}(\mathbf{shifted},\mathbf{red})$
    \Comment{reduce modulo $P(X)$}

\State $\mathbf{y} \gets \mathsf{VAND}(\mathbf{y},\mathsf{VSPLAT}(0xff))$
    \Comment{keep $8$ bits}

\State \Return $\mathbf{y}$
\end{algorithmic}
\end{algorithm}

\begin{algorithm}[H]
\caption{($\mathsf{SCALAR\_VEC}$) Vectorized scalar-by-vector multiplication over $\mathbb{F}_{2^8}$}
\label{alg:gf-mul-scalar-vec-hvx}
\begin{algorithmic}[1]
\Require Scalar field element $a\in\mathbb{F}_{2^8}$
\Require Vector $\mathbf{b}=(b_0,\ldots,b_{63})$ of field elements
\Ensure Vector $(ab_0,\ldots,ab_{63})$

\State $\mathbf{acc} \gets \mathsf{VZERO}()$
    \Comment{product accumulator}
\State $\mathbf{a} \gets \mathsf{VSPLAT}(a)$
    \Comment{broadcast scalar}
\State $\mathbf{one} \gets \mathsf{VSPLAT}(1)$
\State $\mathbf{zero} \gets \mathsf{VZERO}()$

\For{$t \gets 7$ \textbf{downto} $0$}
    \State $\mathbf{b}_t \gets
        \mathsf{VAND}(\mathsf{VSHR}_{t}(\mathbf{b}),\mathbf{one})$
        \Comment{extract bit $t$}

    \State $\mathbf{mask} \gets \mathsf{VSUB}(\mathbf{zero},\mathbf{b}_t)$
        \Comment{bit-to-mask}

    \State $\mathbf{acc} \gets \mathsf{XTIME\_VEC}(\mathbf{acc})$
        \Comment{shift partial product}

    \State $\mathbf{acc} \gets
        \mathsf{VXOR}\bigl(\mathbf{acc},
        \mathsf{VAND}(\mathbf{a},\mathbf{mask})\bigr)$
        \Comment{conditional add}
\EndFor

\State \Return $\mathbf{acc}$
\end{algorithmic}
\end{algorithm}

\begin{algorithm}[H]
\caption{HVX-based vectorized syndrome computation.}
\label{alg:rs-syndrome-hvx}
\textbf{Input:} Received word $\mathbf{c} = (c_0, \dots, c_{n_1-1}) \in \mathbb{F}_{2^8}^{n_1}$.

\textbf{Precomputed:} Power table $\mathcal{A} \in \mathbb{F}_{2^8}^{(n_1-1) \times 64}$ with
\[
\mathcal{A}[j-1] = \big(\alpha^{j},\, \alpha^{2j},\, \ldots,\, \alpha^{(2\delta)j},\, 0,\, \ldots,\, 0\big), \quad 1 \le j < n_1.
\]
Equivalently, $\mathcal{A}[j-1][i] = \mathtt{alpha\_ij\_pow}[i][j]$ for $0 \le i < 2\delta$ and $0$ otherwise. The table is built once and reused across all decode invocations.

\textbf{Output:} Syndrome vector $(\widetilde{S}_0, \dots, \widetilde{S}_{2\delta-1})$ with $\widetilde{S}_i = c_0 \oplus \bigoplus_{j=1}^{n_1-1} c_j\,\alpha^{(i+1)j}$.
\begin{algorithmic}[1]
\State $\mathbf{acc} \gets \textsf{VZERO}()$ \Comment{accumulator over all $2\delta$ syndrome lanes}
\For{$j = 1$ \textbf{to} $n_1 - 1$}
    \State $\mathbf{a}_j \gets \mathcal{A}[j-1]$ \Comment{aligned 64-lane load}
    \State $\mathbf{p} \gets \textsf{SCALAR\_VEC}(c_j, \mathbf{a}_j)$ \Comment{Algorithm 4}
    \State $\mathbf{acc} \gets \textsf{VXOR}(\mathbf{acc}, \mathbf{p})$
\EndFor
\State $\mathbf{acc} \gets \textsf{VXOR}(\mathbf{acc}, \textsf{VSPLAT}(c_0))$ \Comment{inject the constant term}
\State Copy lanes $(\mathbf{acc}[0], \dots, \mathbf{acc}[2\delta-1])$ to $(\widetilde{S}_0, \dots, \widetilde{S}_{2\delta-1})$
\State \Return $(\widetilde{S}_0, \dots, \widetilde{S}_{2\delta-1})$
\end{algorithmic}
\end{algorithm}

\begin{algorithm}
\caption{Branchy Berlekamp--Massey error-locator computation (fastest backend).}
\label{alg:elp}
\begin{algorithmic}[1]
\Require Syndrome vector $\mathbf{S} = (S_0, \dots, S_{2\delta-1})$
\Ensure Error-locator polynomial $\sigma = (\sigma_0, \dots, \sigma_\delta)$ and its degree $\deg\sigma$
\State $\sigma \gets (1, 0, \dots, 0)$;\quad $b \gets (1, 0, \dots, 0)$ \Comment{locator and auxiliary polynomial}
\State $\deg\sigma \gets 0$;\quad $\deg b \gets 0$;\quad $m \gets 1$;\quad $d_\rho \gets 1$
\For{$\mu = 0$ \textbf{to} $2\delta - 1$}
    \State $d \gets S_\mu$
    \For{$i = 1$ \textbf{to} $\deg\sigma$}
        \State $d \gets d \oplus \textsf{GFmul}(\sigma_i, S_{\mu-i})$ \Comment{discrepancy (table-driven GF)}
    \EndFor
    \If{$d \neq 0$}
        \State $\gamma \gets \textsf{GFmul}\!\big(d, \textsf{GFinv}(d_\rho)\big)$
        \State $t \gets \sigma$ \Comment{save current locator}
        \State $u \gets \min(m + \deg b,\ \delta)$
        \For{$i = m$ \textbf{to} $u$}
            \State $\sigma_i \gets \sigma_i \oplus \textsf{GFmul}(\gamma, b_{i-m})$ \Comment{locator update}
        \EndFor
        \If{$2\deg\sigma \le \mu$} \Comment{length change}
            \State $\deg' \gets \mu + 1 - \deg\sigma$
            \State $b \gets t$;\quad $\deg b \gets \deg\sigma$
            \State $\deg\sigma \gets \deg'$;\quad $m \gets 1$;\quad $d_\rho \gets d$
        \Else
            \State $m \gets m + 1$
        \EndIf
    \Else
        \State $m \gets m + 1$
    \EndIf
\EndFor
\State \Return $(\sigma[0..\delta],\ \deg\sigma)$
\end{algorithmic}
\end{algorithm}

\clearpage

\subsection{Abstraction mapping table}
\label{tab:hvx_mapping}

\newcolumntype{C}[1]{>{\centering\arraybackslash}m{#1}}
\newcolumntype{L}{>{\raggedright\arraybackslash}X}
\begin{table}[h]
\centering

\small
\renewcommand{\arraystretch}{1.25}

\begin{tabularx}{\linewidth}{|C{0.20\linewidth}|C{0.36\linewidth}|L|}

\hline
\textbf{Abstraction}
&
\textbf{HVX intrinsic realization}
&
\multicolumn{1}{c|}{\textbf{Technical meaning}}
\\
\hline

\(\mathsf{VADD}\)
&
\texttt{Q6\_Vh\_vadd\_VhVh}
&
Add two vectors lane by lane.
\\
\hline

\(\mathsf{VSUB}\)
&
\texttt{Q6\_Vh\_vsub\_VhVh}
&
Subtract two vectors lane by lane.
\\
\hline

\(\mathsf{VABS}\)
&
\texttt{Q6\_Vh\_vabs\_Vh}
&
Take the absolute value of each signed halfword lane.
\\
\hline

\(\mathsf{VMAX}\)
&
\texttt{Q6\_Vh\_vmax\_VhVh}
&
Compute the lane-wise maximum of two vectors.
\\
\hline

\(\mathsf{VMIN}\)
&
\texttt{Q6\_Vh\_vmin\_VhVh}
&
Compute the lane-wise minimum of two vectors.
\\
\hline

\(\mathsf{VSPLAT}\)
&
\texttt{Q6\_Vh\_vsplat\_R}
&
Broadcast one scalar value to all vector lanes.
\\
\hline

\(\mathsf{VXOR}\)
&
\texttt{Q6\_V\_vxor\_VV}
&
Compute bitwise XOR between two vectors.
\\
\hline

\(\mathsf{VAND}\)
&
\texttt{Q6\_V\_vand\_VV}
&
Compute bitwise AND between two vectors.
\\
\hline

\(\mathsf{VSHL}_{s}\)
&
\texttt{Q6\_Vh\_vasl\_VhR}
&
Shift each lane left by \(s\) bits.
\\
\hline

\(\mathsf{VSHR}_{s}\)
&
\texttt{Q6\_Vuh\_vlsr\_VuhR}
&
Shift each unsigned lane right by \(s\) bits.
\\
\hline

\(\mathsf{VROT}_{s}\)
&
\texttt{Q6\_V\_vror\_VR}
&
Rotate the whole vector register by \(s\) bytes.
\\
\hline

\(\mathsf{VDEAL2}\)
&
\texttt{Q6\_W\_vdeal\_VVR}
&
Deinterleave two input vectors into two reordered output vectors.
\\
\hline

\(\mathsf{VDEALH}\)
&
\texttt{Q6\_Vh\_vdeal\_Vh}
&
Deinterleave halfword lanes inside one vector.
\\
\hline

\(\mathsf{VCMPEQ}\)
&
\texttt{Q6\_Q\_vcmp\_eq\_VhVh}
&
Compare two vectors lane by lane and produce a predicate mask.
\\
\hline

\(\mathsf{VSELECT}\)
&
\texttt{Q6\_V\_vmux\_QVV}
&
Select lanes from two vectors according to a predicate mask.
\\
\hline

\(\mathsf{VREDUCE\_MAX}\)
&
\begin{tabular}{@{}c@{}}
\texttt{Q6\_Vh\_vmax\_VhVh}\\
\texttt{with Q6\_V\_vror\_VR}
\end{tabular}
&
Reduce all lanes to one maximum value using rotations and lane-wise maxima.
\\
\hline

\(\mathsf{VREDUCE\_MIN}\)
&
\begin{tabular}{@{}c@{}}
\texttt{Q6\_Vh\_vmin\_VhVh}\\
\texttt{with Q6\_V\_vror\_VR}
\end{tabular}
&
Reduce all lanes to one minimum value using rotations and lane-wise minima.
\\
\hline

\(\mathsf{VZERO}\) & \texttt{Q6\_V\_Zero} & Generate the zero vector\\
\hline

\end{tabularx}
\vspace{0.5em}
\caption{Mapping between the NPU abstraction and representative Qualcomm Hexagon HVX intrinsics.}
\end{table}

\end{document}